\documentclass[10pt,journal,compsoc]{IEEEtran}
\pdfoutput=1
\usepackage[margin=0.48in]{geometry}

\usepackage{color} 
\usepackage{xcolor,colortbl}
\definecolor{DarkRed}{rgb}{0.8,0,0}
\definecolor{Blue}{rgb}{0,0,0.8}
\definecolor{Gray}{rgb}{0.9,0.9,0.9}

\usepackage{enumitem} 

\newenvironment{EnumerateRoman}{\begin{enumerate}%
\setlength{\parsep}{0pt}\setlength{\topsep}{0.1\baselineskip}\setlength{\itemsep}{0.1\baselineskip}}{\end{enumerate}}

\newenvironment{EnumerateAlpha}{\begin{enumerate}%
\setlength{\parsep}{0pt}\setlength{\topsep}{0.1\baselineskip}\setlength{\itemsep}{0.1\baselineskip}}{\end{enumerate}}

\usepackage{amsmath,amsfonts,amssymb}
\usepackage{algorithmic}
\usepackage[ruled,linesnumbered]{algorithm2e}
\makeatletter
\def\BState{\State\hskip-\ALG@thistlm}
\makeatother
\allowdisplaybreaks
\usepackage{mathrsfs} 
\usepackage{units}
\usepackage{multirow}
\usepackage{tabularx}

\usepackage{lscape}

\usepackage{graphicx} 
\usepackage[tight,footnotesize]{subfigure} 
\usepackage{array} 
\usepackage[footnotesize]{caption} 
\usepackage{tabularx}
\usepackage{ctable}
\usepackage{booktabs}

\usepackage{url}

\usepackage{breakurl}

\newcommand{\Mod}[1]{\ (\text{mod}\ #1)}
\newcommand{\oT}{\textnormal{T}}                                   

\DeclareMathOperator{\rank}{rank}                                      
\DeclareMathOperator{\deter}{det}                                      
\DeclareMathOperator{\sgn}{sgn}                                         
\newcommand{\cL}{\mathcal{L}}                                           
\newcommand{\cR}{\mathcal{R}}                                          
\newcommand{\sX}{\mathscr{X}}                                           

\newcommand{\x}{\delta}
\newcommand{\fp}{f} 
\newcommand{\ad}{d} 
\newcommand{\mumax}{\mu_{\textnormal{max}}}
\newcommand{\cA}{D}                                           
\newcommand{\cW}{W}                                           
\newcommand{\cS}{\mathcal{S}}                                           
\newcommand{\cardS}{\left\vert{S(k,j)}\right\vert}

\newtheorem{proposition}{Proposition}
\newtheorem{definition}[proposition]{Definition}
\newtheorem{theorem}[proposition]{Theorem}
\newtheorem{lemma}[proposition]{Lemma}
\newtheorem{corollary}[proposition]{Corollary}

\newcommand{\BeginProof}{\noindent\hspace{2em}{\itshape Proof: }}
\newcommand{\ProofSquare}{\blacksquare}
\newcommand{\EndProof}{\hspace*{\fill}~$\ProofSquare$\par\unskip}

\begin{document}

\title{Optimized, Direct Sale of Privacy in\\Personal-Data Marketplaces}

\author{Javier Parra-Arnau
\IEEEcompsocitemizethanks{\IEEEcompsocthanksitem The authors is with the Dept. of Comput. Sci., Math., Universitat Rovira i Virgili (URV), Tarragona, Spain.
E-mail: javier.parra@urv.cat}
\thanks{}}

\IEEEtitleabstractindextext{%
\begin{abstract}
Very recently, we are witnessing the emergence of a number of start-ups that enables individuals to sell their private data directly to brokers and businesses.
While this new paradigm may shift the balance of power between individuals and companies that harvest data,
it raises some practical, fundamental questions for users of these services:
how they should decide which data must be vended and which data protected, and what a good deal is.
In this work, we investigate a mechanism that aims at helping users address these questions.
The investigated mechanism relies on a hard-privacy model and allows users
to share partial or complete profile data with broker companies in exchange for an economic reward.
The theoretical analysis of the trade-off between privacy and money posed by such mechanism is the object of this work.
We adopt a generic measure of privacy although part of our analysis focuses on some important examples of Bregman divergences.
We find a parametric solution to the problem of optimal exchange of privacy for money,
and obtain a closed-form expression and characterize the trade-off between profile-disclosure risk and economic reward for several interesting cases.

\end{abstract}

}

\maketitle

\IEEEdisplaynontitleabstractindextext

%
\IEEEpeerreviewmaketitle

\setcounter{footnote}{0} 

\IEEEraisesectionheading{\section{Introduction}\label{sec:Introduction}}
\noindent
\IEEEPARstart{O}{ver} the last recent years, much attention has been paid to government surveillance, and the indiscriminate collection and storage of tremendous amounts of information in the name of national security. However, what most people are not aware of is that a more serious and subtle threat to their privacy is posed by hundreds of companies they have probably never heard of, in the name of commerce.

They are called \emph{data brokers}, and they gather, analyze and package massive amounts of sensitive personal information, which they sell as a product to each other, to advertising companies or marketers, often without our knowledge or consent. A substantial chunk of this is the kind of harmless consumer marketing that has been going on for years. Nevertheless, what has recently changed is the amount and nature of the data being extracted from the Internet and the rapid growth of a tremendously profitable industry that operates with no control whatsoever. Our habits, preferences, our friends, personal data such as date of birth, number of children or home address, and even our daily movements, are some examples of the personal information we are giving up without being aware it is being collected, stored and finally sold to a wide range of companies.

A majority of the population understands that this is part of an unwritten contract whereby
they get content and services free in return for letting advertisers track their behavior; this is the barker economy that, for example, currently sustains the Web.
But while a significant part of the population finds this tracking invasive,
there are people who do not give a toss about
being mined for data~\cite{AdblockPlus15Survey}. 

Very recently we are witnessing the emergence of a number of start-ups that hope to exploit this by buying access to our social-networks accounts and banking data.
\begin{figure}[t!]
\centering
\includegraphics[width=\columnwidth]{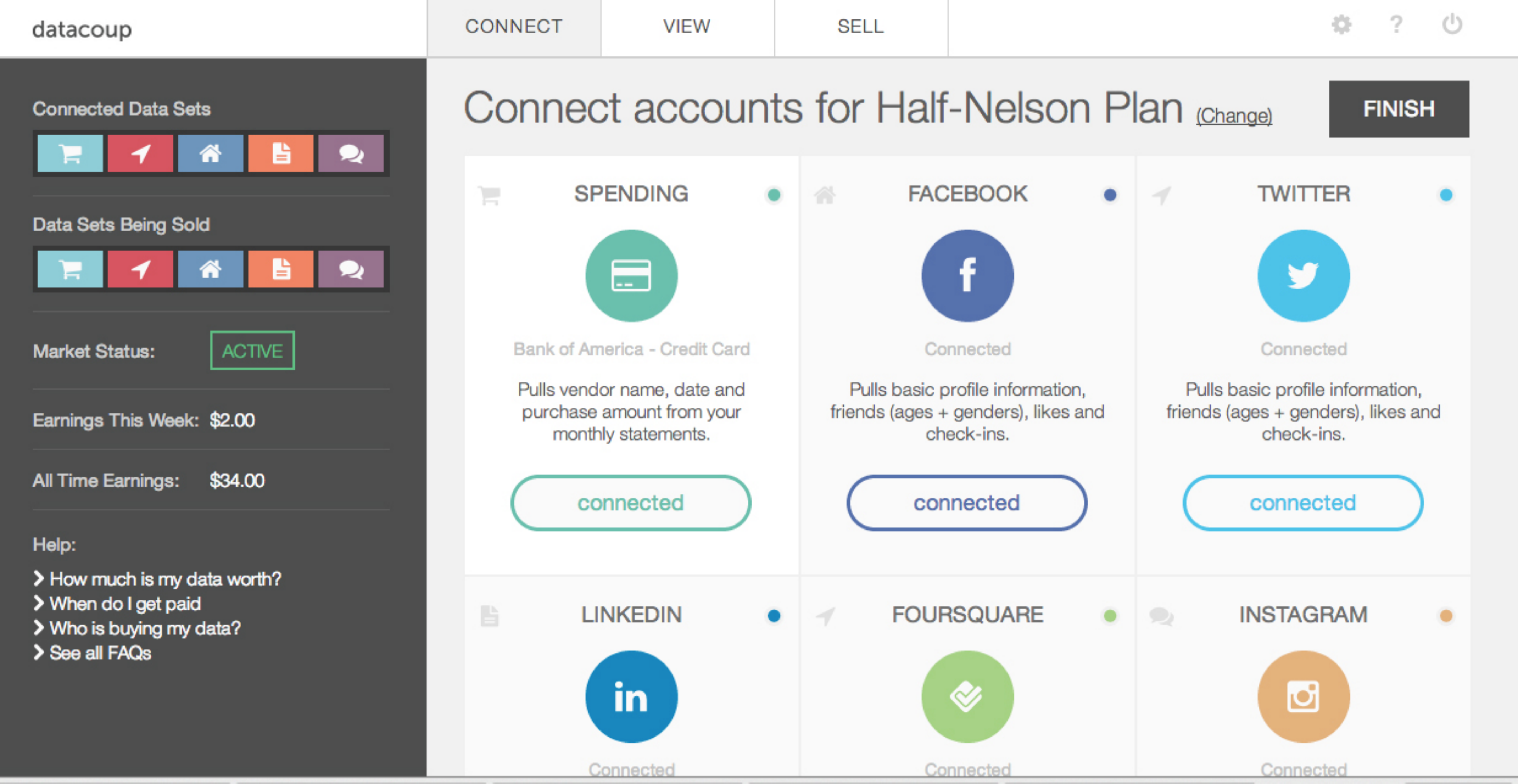}
\caption{Screenshot of Datacoup which allows users to earn money by sharing their personal data.}
\label{fig:conceptualtradeoff}
\end{figure}
One such company is Datacoup, which lets users connect their apps and services via APIs in order to sell their data.
Datacoup and similar start-ups, however, do not provide raw data to potential purchasers, among others, retailers, insurance companies and banks.
Rather, they typically build a profile that gives these companies an overview of a user's data.

The emergence of these start-ups is expected to provide a win-win situation both for users and data buyers.
On the one hand, users will receive payments, discounts or various rewards from purchasing companies,
which will take advantage of the notion that users are receiving
a poor deal when they trade personal data in for access to ``free'' services. 
On the other hand,
companies will earn more money as the quality of the data these start-ups will offer to them will be much greater than that currently provided by traditional brokers ---the problem with the current brokers is often the stale and inaccurate data~\cite{Sile15CBNC}.

The possibility that individuals may vend their private data \emph{directly} to businesses and retailers will be one step closer with the emergence of companies like Datacoup. 
For many, this can have a liberating effect. It permeates the opaque data-exchange process with a new transparency, and empowers online users
to decide what to sell and what to retain. 
However, the prospect of people selling data directly to brokers poses a myriad of new problems for their owners. How should they manage the sale of their data? How should they decide which elements must be offered up and which data protected? What is a good deal?

\subsection{Contribution and Plan of this Paper}
\label{sec:Introduction:Contribution}
\noindent
In this paper, we investigate a mechanism that aims at helping users address these questions.
The investigated mechanism builds upon the new data-purchasing paradigm developed by broker companies like Datacoup, CitizenMe and DataWallet, which allows users to sell their private data \emph{directly} to businesses and retailers.
The mechanism analyzed in this work, however, relies on a variant of such paradigm which gives priority to users, in the sense that they are willing to disclose partial or complete profile data only when they have an offer on the table from a data buyer, and not the other way round.
Also, we assume a hard-privacy model by which users take charge of protecting their private data on their own,
without the requirement of trusted intermediaries.

The theoretical analysis of the trade-off between disclosure risk and economic reward posed by said mechanism is the object of this work.
We tackle the issue in a mathematically, systematic fashion,
drawing upon the methodology of multiobjective optimization.
We present a mathematical formulation of optimal exchange of profile data for money, which takes into account the trade-off between both aspects by treating them as two sides of the same coin, and which contemplates a rich variety of functions as quantifiable measures of user-profile privacy.
Our theoretical analysis finds a closed-form solution to the problem of optimal sale of profile data, and characterizes the optimal trade-off between privacy and money.

Sec.~\ref{sec:Mechanism} introduces our mechanism for the exchange of profile data for money, proposes a model of user profile, and formulates the trade-off between privacy and economic reward.
We proceed with a theoretical analysis in Sec.~\ref{sec:Theory}, while Sec.~\ref{sec:Examples} numerically illustrates the main results.
Next, Sec.~\ref{sec:SotA} reviews the state of art and conclusions are drawn in Sec.~\ref{sec:Conclusions}.
\section{A Mechanism for the Exchange of Private Data for Money}
\label{sec:Mechanism}
\noindent
In this section, we present a mechanism that allows users to share portions of their profile with data-broker companies, in exchange for an economic reward.
The description of our mechanism is prefaced by a brief introduction of the concept of hard privacy and our data-purchasing model.

\begin{figure}[tb]
\centering
\includegraphics[scale=0.55]{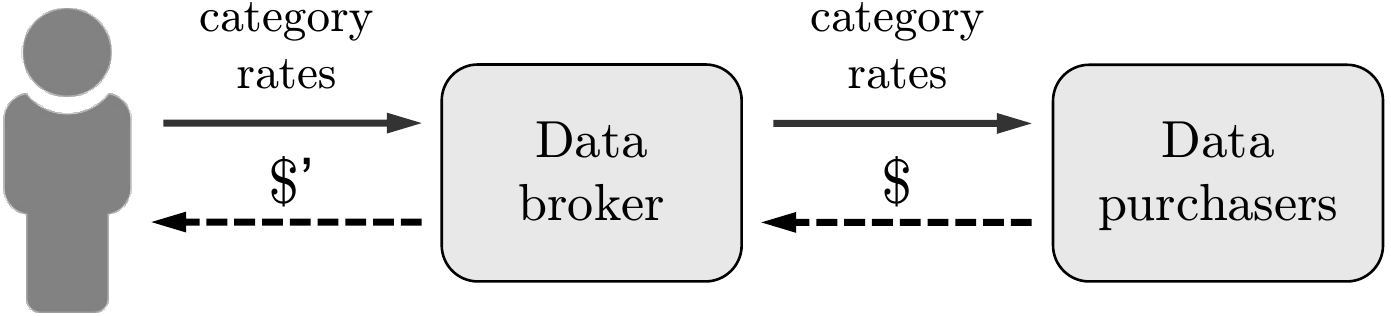}
\caption{Conceptual depiction of the data-purchasing model assumed in this work. In this model, users first send the data broker their category rates, that is, the money they would like to be paid for completely exposing their actual interests in each of the categories of a profile. Based on the rates chosen for each category, data buyers decide then whether to pay the user for learning their profile and gaining access to the underlying data. Finally, depending on the offer made, the disclosure may range from portions of their profile to the complete actual profile.}
\label{fig:UserProfile}
\end{figure}

\subsection{Hard-Privacy and Data-Purchasing Model}
\label{sec:Mechanism:HardSoft}
\noindent
Privacy-enhancing technologies (PETs) can be classified depending on the level of trust placed by their users~\cite{Danezis07Talk,Mina10PHD}.
A privacy mechanism providing \emph{soft privacy} assumes that users entrust their private data to an entity,
which is thereafter responsible for the protection of their data.
In the literature, numerous attempts to protect privacy have followed the traditional method of
pseudonymization and anonymization~\cite{Chaum81CACM},
which are essentially based on the assumptions of soft privacy.
Unfortunately, these methods are not completely
effective. 
they normally come at the cost of infrastructure, and suppose that users are willing to trust other parties.

The mechanism investigated in this work, per contra, capitalizes on the principle of \emph{hard privacy},
which assumes that users mistrust communicating entities and are therefore reluctant to delegate the protection of their privacy to them.
In the motivating scenario of this work, hard privacy means that users do not trust the new data brokerage firms ---not to mention data purchasers--- to safeguard their personal data.
Consequently, because users just trust themselves, it is their own responsibility to protect their privacy.

In the data-purchasing model supported by most of these new data brokers,
users, just after registering ---and without having received any money yet---, must give these companies access to one or several of their accounts.
As mentioned in the introductory section, brokers at first do not provide raw data to potential buyers.
Rather, purchasers are shown a \emph{profile} of the data available at those accounts,
which gives them an accurate-enough description of a user's interests, 
so as to make a decision on whether to bid or not for that particular user.
If a purchaser is finally interested in a given profile, the data of the corresponding account are sold at the price fixed by the broker.
Obviously, the buyer can at that point verify that the purchased data corresponds to the profile it was initially shown, that is, it can check the profile was built from such data.
At the end of this process, users are notified of the purchase.

In this work, we assume a variation of this data-purchasing model that reverses the order in which transactions are made. In essence, we consider a scenario where, first, users receive an economic reward, and then, based on that reward, their data are partly or completely disclosed to the bidding companies; this variation is in line with the literature of pricing private data~\cite{Roth12SIGECOM}, examined in Sec.~\ref{sec:SotA}.
Also, we contemplate that users themselves take charge of this information disclosure, without the intervention of any external entity, following the principle of hard privacy.

More specifically, users of our data-buying model first notify brokers of the compensation they wish to receive for fully disclosing each of the components of their profile ---we shall henceforth refer to these compensations as \emph{category rates}.
For example, if profiles represent purchasing habits across a number of categories,
a user might specify low rates for completely revealing their shopping activity in groceries, and they might impose higher prices on more sensitive purchasing categories like health care.
Afterwards, based on these rates, interested buyers try to make a bid for the entire profile.
However, as commented above, it is now up to the user to decide whether to accept or decline the offer.
Should it be accepted, the user would disclose their profile according to the money offered, and give the buyer ---and the intermediary broker--- access to the corresponding data.

As we shall describe more precisely in the coming subsections, we shall assume a controlled disclosure of user information that will hinge upon the particular economic reward given. Basically, the more money is offered to a user, the more similar the disclosed profile will be to the actual one.
Furthermore, we shall assume that there exists a communication protocol enabling this exchange of information for money,
and that users behave honestly in \emph{all} steps of said data-transaction process.
This work does not tackle the practical details of an implementation of this protocol and the buying model described above.
This is nevertheless an important issue, and dispelling the assumption that must behave honestly is one of the many exciting directions for future work.

\subsection{User-Profile Representation}
\label{sec:Mechanism:Privacy}
\noindent
We model user private data (e.g., posts and tags on social networks, transactions in a bank account) as a sequence of random variables (r.v.'s) taking on values in a common finite alphabet of categories, in particular the set $\sX = \{1,\ldots,n\}$ for some integer $n\geqslant2$.
In our mathematical model, we assume these r.v.'s are independent and identically distributed.
This assumption allows us to represent the profile of a user by means of the probability mass function (PMF) according to which such r.v.'s are distributed, a model that is widely accepted in the privacy literature~\cite{Xu07WWW,Toubiana10SNDSS,Fredrikson11SP}.

Conceptually, we may interpret a profile as a histogram of relative frequencies of user data within that set of categories.
For instance, in the case of a bank account, grocery shopping and traveling expenses could be two categories.
In the case of social-networks accounts, on the other hand, posts could be classified across topics such as politics, sports and technology.

In our scenario of data monetization,
users may accept unveiling some pieces of their profile, in exchange for an economic reward.
Users may consider, for example, revealing a fraction of their purchases on Zappoos, and may avoid disclosing their payments at nightclubs.
Clearly, depending on the offered compensation,
the profile observed by the broker and buying companies will resemble, to a greater or lesser extent, the genuine shopping habits of the user.
In this work, we shall refer to these two profiles as the \emph{actual user profile} and the \emph{apparent user profile},
and denote them by $q$ and $t$, respectively.

\subsection{Privacy Models}
\label{sec:Mechanism:Privacy}
\noindent
Before deciding how to disclose a profile for a given reward,
users must bear in mind the privacy objective they want to achieve by such disclosure.
In the literature of information privacy, this objective is inextricably linked to the concrete assumptions about the attacker against which a user wants to protect. This is known as the \emph{adversary model} and its importance lies in the fact that the level of privacy provided is measured with respect to it.

In this work, we consider two privacy objectives for users, which may also be interpreted from an attacker perspective;
in our case, data brokers, data-buying companies and in general any entity with access to profile information may all be regarded as privacy adversaries.
\begin{itemize}
\item On the one hand, we assume a \emph{profile-density} model, in which a user wishes to make their profile more common, trying to hide it in the crowd.
\item On the other hand, we consider a \emph{classification} model where the user does not want to be identified as a member of a given group of users.
\end{itemize}

In terms of an adversary model, the former objective could be defined under the assumption that the attacker aims at targeting peculiar users, that is, users who deviate from a typical behavior. The latter model, on the other hand, could fit with an adversary who wishes to label a user as belonging to a particular group. In either case, the ultimate aim of an attacker could be from price discrimination to social sorting.

In our mathematical model, the selection of either privacy model entails choosing a reference, \emph{initial profile} $p$ the user wishes to impersonate when no money is offered for their data. For example, in the profile-density model, a user might want to exhibit very common interests or habits and $p$ might therefore be the average profile of the population. In the classification model, the user might be comfortable with showing the profile of a less-sensitive group.
As we shall explain in the next subsection, the initial profile will provide a ``neutral'' starting point for the disclosure of the actual profile $q$.

\subsection{Disclosure Mechanism and Privacy Function}
\label{sec:Mechanism:PrivacyFunction}
\noindent
In this section, we propose a profile-disclosure mechanism suitable for the data-buying and privacy models described previously.
The proposed technique operates between these two extreme cases. When there is no economic compensation for having access to a user account,
the disclosed profile coincides with the initial distribution $p$, and the observation of this information by the data broker or potential purchasers does not pose any privacy risk to the user.
When the user is offered sufficient reward, however, the actual profile $q$ is fully disclosed and their privacy completely compromised.

\begin{figure}[tb]
\centering
\includegraphics[scale=0.40]{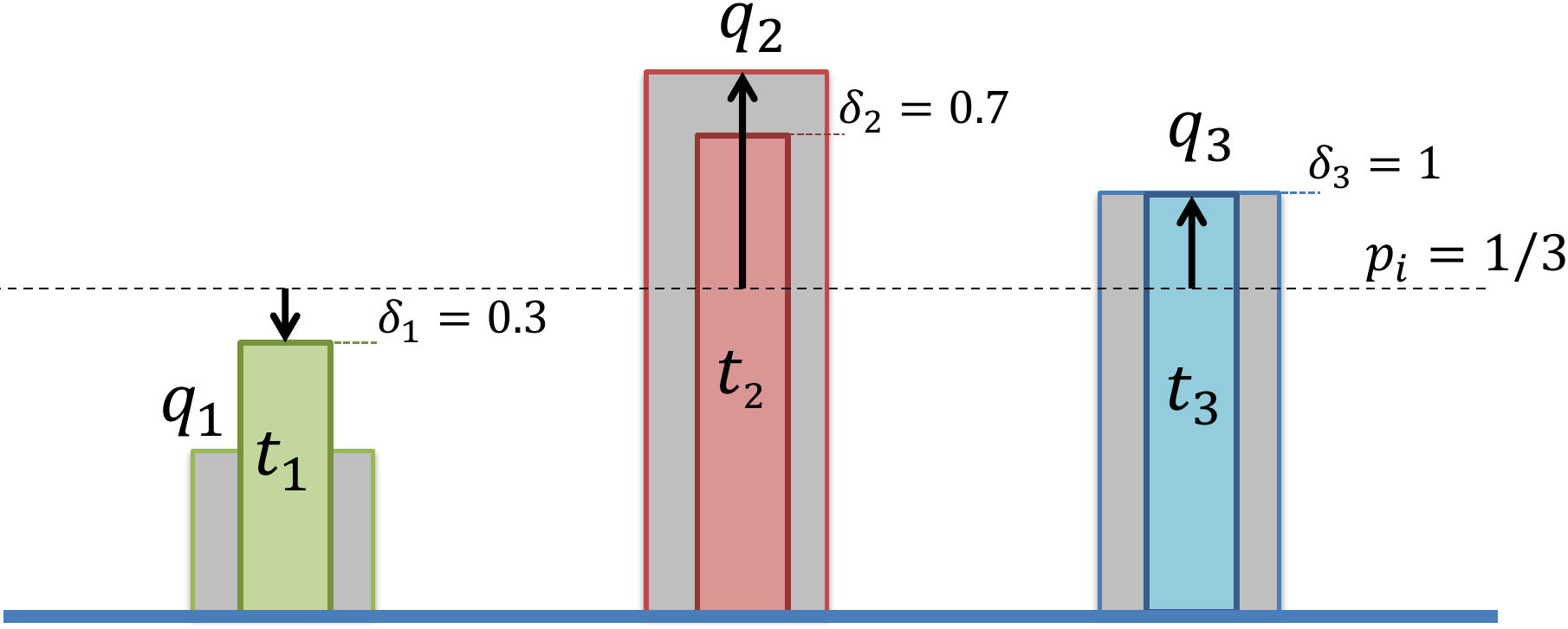}
\caption{We provide an example of how our profile-disclosure mechanism operates. In this example, we consider category rates of 1 dollar for each of the $n=3$ categories, and a 2-dollar offer by a purchasing company. We show an apparent profile $t$ that results from applying a certain disclosure strategy on the actual profile $q$. The disclosure departs from the uniform distribution. The selected strategy fully reveals the interest of the user in the category 3. However, the compensation offered does not allow them to do the same with the categories 1 and 2. Rather, the user decides to expose 30 and 70 percent of the interest values in these two categories respectively, which equates to the received reward.}
\label{fig:DisclosureMechanism}
\end{figure}

Our disclosure mechanism reveals the deviation of the user's initial, false interest to the actual value.
In formal terms, we define the \emph{disclosure rate} $\x_i$ as the percentage of disclosure lying on the line segment between $p_i$ and $q_i$.
Concordantly, we define the user's apparent profile $t$ as the convex combination $t = (1- \x)\,p + \x\,q$, where $\x = (\x_1,\ldots,\x_n)$ is some \emph{disclosure strategy} specified by the user.
The disclosure mechanism may be interpreted intuitively as a roller blind. The starting position $\x=0$ corresponds to leaving the roller in the value $p$, that is, $t=p$. Depending on whether $q_i<p_i$ or $q_i>p_i$, a positive $\x$ may translate into lowering or raising the roller respectively. Fig.~\ref{fig:DisclosureMechanism} illustrates this effect for a uniform initial profile, that is, $p_i=1/n$ for all $i=1,\ldots,n$.

In our model, the user therefore must decide a disclosure strategy that shifts $t$ from the initial PMF to the actual one;
clearly, the disclosed information must equate to the money offered by the data purchaser.
The question that follows naturally is, what is the privacy loss due to this shift, or said otherwise, how do we measure the privacy of the apparent profile?

In this work, we do not contemplate a single, specific privacy metric, nor consider that all users evaluate privacy the same way.
Instead, each user is allowed to choose the most appropriate measure for their privacy requirements.
In particular, we quantify a user's \emph{privacy risk} generically as
\begin{equation*}
\cR = \fp(t,p) = \fp((1- \x)\,p + \x\,q, p),
\end{equation*}
where $\fp\colon (t,p) \mapsto \fp(t,p)$ is a \emph{privacy function}
that measures the extent to which the user is discontent when the initial profile is $p$ and the apparent profile is $t$.

A particularly interesting class of those privacy functions are the \emph{dissimilarity or distance metrics},
which have been extensively used to measure the privacy of user profiles.
The intuitive reasoning behind these metrics is that apparent profiles closer to $p$ offer better privacy protection than those closer to $q$,
which is consistent with the two privacy models described in Sec.~\ref{sec:Mechanism:Privacy}.
Examples of these functions comprise the Euclidean distance, Kullback-Leibler (KL) divergence~\cite{Cover06B}, and the cosine and Hamming distances.

 \subsection{Formulation of the Optimal Trade-Off between Privacy and Money}
 \label{sec:Mechanism:Formulation}
 \noindent
Equipped with a measure of the privacy risk incurred by a disclosure strategy,
the proposed mechanism aims at finding the strategy that yields the minimum risk for a given reward.
Next, we formalize the problem of choosing said strategy as a multiobjective optimization problem whereby
users can configure a suitable trade-off between privacy and money.

Let $w = (w_1, \ldots, w_n)$ be the tuple of category rates specified by a user, that is, the amount of money they require to completely disclose their interests or habits in each category.
Since in our data-buying model users have no motivation for giving their private data for free, we shall assume these rates are positive.
Accordingly, for a given economic compensation $\mu$, we define the \emph{privacy-money function} as
\begin{equation}
\label{eqn:PrivacyMoneyFunction}
\cR(\mu)=  \min_{\substack{\x\\\sum_i w_i \x_i = \mu,\\ \sum_i t_i = 1,\\ 0 \leqslant \x_i \leqslant 1}} \fp(t,p),
\end{equation}
which characterizes the optimal trade-off between privacy and economic compensation.

Conceptually, the result of this optimization problem is a disclosure strategy $\x^*$ that tells us, for a given amount of money, how to unveil a profile so that the level of privacy is maximized.
Intuitively, if $\fp$ is a profile-similarity function, the disclosure is chosen to minimize the discrepancies between the apparent and the initial profiles. Naturally, the minimization must satisfy that the compensation offered is effectively exchanged for private information.
This is what the condition $\mu = \sum_i w_i \x_i$ means.
The other equality condition, $\sum_i t_i = 1$, merely reflects that the resulting apparent profile must be a probability distribution.

In closing, the problem~\eqref{eqn:PrivacyMoneyFunction} gives a disclosure rule that not only assists users in protecting their privacy, but also allows them to find the optimal exchange of privacy for money.

\section{Optimal Disclosure of Profile Information}
\label{sec:Theory}
\noindent
This section is entirely devoted to the theoretical analysis of the privacy-money function~\eqref{eqn:PrivacyMoneyFunction} defined in Sec.~\ref{sec:Mechanism:Formulation}.
In our attempt to characterize the trade-off between privacy risk and money,
we shall present a solution to the optimization problem inherent in the definition of this function.
Afterwards, we shall analyze some fundamental properties of said trade-off for several interesting cases.
For the sake of brevity, our theoretical analysis only contemplates the case when all given probabilities and category rates are strictly positive:
\begin{equation}
\label{eqn:PositivityAssumption}
q_i, p_i >0 \textnormal{ for all }i=1,\dots,n.
\end{equation}

Without loss of generality, we shall assume that
\begin{equation}
\label{eqn:NoEqual}
q_i \neq p_i \textnormal{ for all }i=1,\dots,n.
\end{equation}
We note that we can always restrict the alphabet $\mathscr{X}$ to those categories where $q_i \neq p_i$ holds, and redefine the two probability distributions accordingly.

In this work, we shall limit our analysis to the case of privacy functions $\fp\colon(t,p) \mapsto \fp(t,p)$ that are twice differentiable on the interior of their domains. In addition, we shall consider these functions capture a measure of \emph{dissimilarity or distance} between the PMFs $t$ and $p$, and accordingly assume that $\fp(t,p) \geqslant 0$, with equality if, and only if, $t=p$.
Occasionally, we shall denote $\fp$ more compactly as a function of $\x$, on account of the fact that $t=(1-\x)\,p + \x\,q$, and that $p$ and $q$ are fixed variables.

Before establishing some notational aspects and diving into the mathematical analysis,
it is immediate from the definition of the privacy-money function and the assumptions made above that its initial value is
$\cR(0)=0$.
The characterization of the optimal trade-off curve modeled by $\cR(\mu)$ at any other values of $\mu$ is the focus of this section.

\subsection{Notation and Preliminaries}
\label{sec:Theory:Notation}
\noindent
We shall adopt the same notation for vectors used in~\cite{Boyd04B}.
Specifically, we delimit vectors and matrices with square brackets, with the components separated by space, and use parentheses to construct column vectors from comma separated lists.

Occasionally, we shall use the notation $x^{\oT}y$ to indicate the standard inner product on $\mathbb{R}^n$, $\sum_{i=1}^n x_i y_i·$,
and $\|\cdot\|$ to denote the Euclidean norm, i.e., $\|x\|= (x^{\oT}x)^{1/2}$.
Recall~\cite{Boyd04B} that a \emph{hyperplane} is a set of the form
$$\{x : v^{\oT} x = b\},$$
where $v\in \mathbb{R}^n$, $v\neq 0$, and $b\in \mathbb{R}$. Geometrically, a hyperplane may be regarded as the set of points with a constant inner product to a vector $v$. Note that a hyperplane separates $\mathbb{R}^n$ into two halves; each of these halves is called a \emph{halfspace}.
The results developed in the coming subsections will build upon a particular intersection of halfspaces, usually referred to as \emph{slab}. Concretely, a slab is a set of the form $$\{x : b_l \leqslant v^{\oT} x \leqslant b_u\},$$
the boundary of which are two hyperplanes.
Informally, we shall refer to them as the lower and upper hyperplanes.

\subsection{Monotonicity and Convexity}
\label{sec:Theory:MonoConc}
\noindent
Our first theoretical characterization, namely Theorems~\ref{thr:Monotonicity} and~\ref{thr:Convexity}, investigates two elementary properties of the privacy-money trade-off.
The theorems in question show that the trade-off is nondecreasing and convex.
The importance of these two properties is that they confirm the evidence that an economic reward will never lead to an improvement in privacy protection.
In other words, accepting money from a data purchaser does not lower privacy risk.
Together, these two results will allow us to determine the shape of $\cR(\mu)$.

Before proceeding, define $\mumax = \sum_i w_i$ and note that when $\mu = \mumax$,
the equality condition $\sum_i w_i\x_i  = \mu$ implies $\x_i =1$ for all $i$.
Hence, $\cR(\mumax) = \fp(q,p)$.
Also, observe that the privacy-money function is not defined for a compensation $\mu>\mumax$ since the optimization problem inherent in the definition of this function is not feasible.

\noindent
\begin{theorem}[Monotonicity]
\label{thr:Monotonicity}
The privacy-money function $\cR(\mu)$ is nondecreasing.
\end{theorem}

\BeginProof
Consider an alternative privacy-money function $\cR^{a}(\mu)$ where the condition
$\sum_i w_i  \x_i=\mu$ is replaced by these two inequality constraints, $\mu \leqslant \sum_i w_i  \x_i \leqslant \mumax$.
We shall first show that this function is nondecreasing and, based on it, we shall prove the monotonicity of $\cR(\mu)$.

Let $0\leqslant\mu<\mu'\leqslant \mumax$,
and denote by $\x'$ the solution to the minimization problem corresponding to $\cR^{a}(\mu')$.
Clearly, $\x'$
is feasible to the problem $\cR^{a}(\mu)$ since $\mu'>\mu$.
Because the feasibility of~$\x'$ does not necessarily imply that it is a minimizer of the problem corresponding to $\cR^{a}(\mu)$, it follows that
$$\cR^a(\mu)\leqslant \fp\left((1-\x')p + \x'q,\,p\right) =\cR^a(\mu'),$$
and hence that the alternative privacy-money function is nondecreasing.

This alternative function can be expressed in terms of the original one,
by taking $\cR(\mu)$ as an inner optimization problem of $\cR^a(\mu)$, namely $\cR^a(\mu)=  \min_{\substack{\mu\leqslant \alpha\leqslant \mumax}} \cR(\alpha)$.
Based on this expression, it is straightforward to verify that the only condition consistent with the fact that $\cR^a(\mu)$ is nondecreasing
is that $\cR(\mu)$ be nondecreasing too.
\EndProof

Next, we define an interesting property borrowed from~\cite{Cover06B} for KL divergence,
that will be used in Theorem~\ref{thr:Convexity} to show the convexity of the privacy-money function.
\begin{definition}
\label{def:convexitypair}
A function $\fp(t,p)$ is \emph{convex in the pair} $(t,p)$ if
\begin{multline}
\label{eqn:ConvexPairs}
\fp (\lambda t_1 + (1-\lambda) t_2, \lambda p_1 + (1- \lambda) p_2) \\\leqslant \lambda \fp(t_1,p_1) + (1 - \lambda) \fp(t_2,p_2),
\end{multline}
for all pairs of probability distributions $(t_1,p_1)$ and $(t_2,p_2)$ and all $0\leqslant \lambda \leqslant 1$.
\end{definition}

\noindent
\begin{theorem}[Convexity]
\label{thr:Convexity}
If $\fp(t,p)$ is convex in the pair $(t,p)$, then the corresponding privacy-money function $\cR(\mu)$ is convex.
\end{theorem}

\BeginProof
The proof closely follows the proof of Theorem~1 of~\cite{Rebollo10IT}.
We proceed by checking the definition of convexity, that is, that
$$(1-\lambda)\,\cR(\mu) +\lambda\,\cR(\mu') \geqslant \cR((1-\lambda)\,\mu + \lambda\,\mu')$$
for all $0\leqslant \mu < \mu' \leqslant \mumax$ and all $0\leqslant \lambda \leqslant 1$.
Denote by $\x$ and $\x'$ the solutions to $\cR(\mu)$ and $\cR(\mu')$, respectively,
and define $\x_{\lambda} = (1 - \lambda)\,\x + \lambda\,\x'$.
Accordingly,
\begin{multline*}
\begin{aligned}
  (1-\lambda)\,\cR(\mu) +\lambda\,\cR(\mu')  &=                                          (1-\lambda)\,\fp((1-\x)\,p + \x\,q,\,p)\\
    &\quad +                                    \lambda\,\fp((1-\x')\,p + \x'q,\,p) \\
    &\,\,{\stackrel{\textnormal{(a)}}{\geqslant}}\,   \fp\Big((1-\lambda)\,((1-\x)\,p+\x\,q)\\
    &\quad +                                    \lambda\,((1-\x')\,p+\x'q),\,p\Big)\\
    &=                                          \fp((1-\x_{\lambda})\,p + \x_{\lambda}\,q,\,p)\\
    &\,\,{\stackrel{\textnormal{(b)}}{\geqslant}}\,    \cR((1-\lambda)\,\mu + \lambda\,\mu'),
\end{aligned}
\end{multline*}
where
\begin{EnumerateAlpha}
\item follows from the fact that $\fp(t,p)$ is convex in the pairs of probability distributions~\cite[\S 2]{Cover06B}, and
\item reflects that $\x_\lambda$ is not necessarily the solution to the minimization problem $\cR((1-\lambda)\,\mu + \lambda\,\mu')$.
\EndProof
\end{EnumerateAlpha}

The convexity of the privacy-money function~\eqref{eqn:PrivacyMoneyFunction} guarantees
its continuity on the interior of its domain, namely $(0,\mumax)$.
However, it can be checked,
directly from the definition of $\cR(\mu)$,
that continuity also holds at the interval endpoints, 0 and~$\mumax$.

Lastly, we would like to point out the generality of the results shown in this subsection,
which are valid for a wide variety of privacy functions $\fp(t,p)$, provided that they are non-negative, twice differentiable and convex in the pair $(t,p)$. Some examples of functions meeting these properties are the squared Euclidean distance (SED) and KL divergence.

\subsection{Parametric Solution}
\label{sec:Theory:ResourceAllocation}
\noindent
Our next result, Lemma~\ref{lem:genericproblem}, provides a parametric solution to the minimization problem involved in the formulation of the privacy-money trade-off~\eqref{eqn:PrivacyMoneyFunction} for certain privacy functions.
Even though said lemma provides a parametric-form solution,
fortunately we shall be able to proceed towards an explicit closed-form expression, albeit piecewise, for some special cases and values of $n$.
For the sake of notational compactness, we define the difference tuple $\ad = \left(q_1-p_1,\dots,q_n-p_n\right)$.

\begin{lemma}[General Parametric Solution]\label{lem:genericproblem}
Let $\fp$ be additively separable into the functions $\fp_i$ for $i=1,\dots,n$.
For all $i$, let $\fp_i:[0,1]\to\mathbb{R}$ be twice differentiable in the interior of its domain, with $\fp''_i>0$, and hence strictly convex.
Because $\fp''_i>0$, $f'_i$ is strictly increasing and therefore invertible.
Denote the inverse by ${\fp'_i}^{-1}$.
Now consider the following optimization problem in the variables $\x_1,\dots,\x_n$:
\begin{align}
\label{eqn:actualproblem}
&\textnormal{minimize }     &&\sum_{i=1}^n \fp_i(\x_i) \nonumber \\
&\textnormal{subject to }    &&0\leqslant \x_i\leqslant 1 \textnormal{ for } i =1,\ldots,n,\\
                                                &&&\sum_{i=1}^n \ad_i \x_i  = 0 \textnormal{ and } \sum_{i=1}^n w_i \x_i = \mu. \nonumber 
\end{align}
The solution to the problem exists, is unique and of the form
\begin{equation*}
\x^*_i=\max \left\{0,\min\{{\fp'_i}^{-1}(\alpha\,\ad_i + \beta\,w_i),1\} \right\},
\end{equation*}
for some real numbers $\alpha,\beta$ such that $\sum_i \ad_i \x^*_i=0$ and $\sum_i w_i \x^*_i = \mu$.
\end{lemma}

\BeginProof
We organize the proof in two steps.
In the first step, we show that the optimization problem stated in the lemma is convex; 
then we apply Karush-Kuhn-Tucker (KKT) conditions to said problem, and finally reformulate these conditions into a reduced number of equations.
The bulk of this proof comes later, in the second step, where we proceed to solve the system of equations.

To see that the problem is convex, simply observe that the objective function $\fp$ is the sum of strictly convex functions $\fp_i$, and
that the inequality and equality constraint functions are affine.
The existence and uniqueness of the solution is then a consequence of the fact that we minimize a strictly convex function over a convex set.
Since the objective and constraint functions are also differentiable and Slater's constraint qualification holds, KKT conditions are necessary and sufficient conditions for optimality~\cite[\S 5]{Boyd04B}.
The application of these optimality conditions leads to the following Lagrangian cost,
\begin{multline*}
\cL =\sum \fp_i(\x_i) - \sum \lambda_i \x_i \\
+\sum \mu_i (\x_i - 1) - \alpha \sum \ad_i \x_i - \beta\left(\sum w_i \x_i - \mu\right),
\end{multline*}
and finally to the conditions
\begin{equation*}
\setlength\arraycolsep{-1.29em}
\begin{array}{lrr}
\rule{0pt}{1ex}
\fp'_i(\x_i)-\lambda_i+\mu_i - \alpha \ad_i - \beta w_i =0      &\,\, \textnormal{(dual optimality),}\\
\rule{0pt}{3ex}
\lambda_i \x_i = 0,\, \mu_i(\x_i - 1)=0               & \textnormal{(complementary slackness),}\\
\rule{0pt}{3ex}
\lambda_i, \mu_i \geqslant 0            &\,\, \textnormal{(dual feasibility),}\\
\rule{0pt}{3ex}
0\leqslant \x_i\leqslant 1,\, \sum \ad_i\x_i=0,\,\sum w_i\x_i=\mu       &\,\, \textnormal{(primal feasibility).} \\
\end{array}
\end{equation*}

We may rewrite the dual optimality condition as
$\lambda_i = \fp'_i(\x_i) +\mu_i - \alpha \ad_i - \beta w_i$
and $\mu_i = \alpha \ad_i + \beta w_i - \fp'_i(\x_i) +\lambda_i$.
By eliminating the slack variables $\lambda_i,\,\mu_i$, and
by substituting the above expressions into the complementary slackness conditions,
we can formulate the dual optimality and complementary slackness conditions equivalently as
\begin{align}
        &\fp'_i(\x_i) + \mu_i \geqslant \, \alpha \ad_i +\beta w_i, \label{eqn:Lemma1}\\
        &\fp'_i(\x_i) - \lambda_i \leqslant \, \alpha \ad_i + \beta w_i, \label{eqn:Lemma2}\\
         &(\fp'_i(\x_i) + \mu_i - \alpha \ad_i - \beta w_i)\,\x_i=0, \label{eqn:Lemma3}\\
        &(\fp'_i(\x_i) - \lambda_i - \alpha \ad_i - \beta w_i)\, (\x_i - 1)=0. \label{eqn:Lemma4}
\end{align}
In the following, we shall proceed to solve these equations which, together with the primal and dual feasibility conditions, are necessary and sufficient conditions for optimality.
To this end, we consider these three possibilities for each~$i$: $\x_i = 0$, $0 < \x_i < 1$ and $\x_i = 1$.

We first assume $\x_i = 0$.
By complementary slackness, it follows that $\mu_i = 0$ and, in virtue of~\eqref{eqn:Lemma1}, that $\fp'_i(0) \geqslant \alpha \ad_i +\beta w_i.$
We now suppose that this latter inequality holds and that $\x_i >0$.
However, if $\x_i$ is positive, by equation~\eqref{eqn:Lemma2} we have $\fp'_i(\x_i) \leqslant \alpha \ad_i +\beta w_i$,
which contradicts the fact that $\fp'_i$ is strictly increasing.
Hence, $\x_i = 0$ if, and only if, $\alpha \ad_i + \beta w_i \leqslant \fp'_i(0)$.

Next, we consider the case $0 < \x_i < 1$.
Note that, when $\x_i >0$, it follows from the conditions~\eqref{eqn:Lemma2} and~\eqref{eqn:Lemma3} that $\fp'_i(\x_i) \leqslant \alpha \ad_i +\beta w_i$, which, by the strict monotonicity of $\fp'_i$, implies $\fp'_i(0) < \alpha \ad_i +\beta w_i$.
On the other hand, when $\x_i <1$, the conditions~\eqref{eqn:Lemma4} and~\eqref{eqn:Lemma1} and again the fact that $\fp'_i$ is strictly increasing imply that $\alpha \ad_i +\beta w_i < \fp'_i(1)$.

To show the converse, that is, that $\fp'_i(0) < \alpha \ad_i +\beta w_i < \fp'_i(1)$ is a sufficient condition for $0 < \x_i < 1$, we proceed by contradiction and suppose that the left-hand side inequality holds 
and the solution is zero.
Under this assumption, equation~\eqref{eqn:Lemma4} implies that $\mu_i=0$, and in turn that $\fp'_i(0) \geqslant \alpha \ad_i + \beta w_i$,
which is inconsistent with the fact that $\fp'_i$ is strictly increasing.
Further, assuming $\alpha \ad_i +\beta w_i < \fp'_i(1)$ and $\x_i=1$ implies that $\lambda_i=0$ and, on account of~\eqref{eqn:Lemma2}, that $\fp'_i(1) \leqslant \alpha \ad_i +\beta w_i$, a contradiction.
Consequently, the condition $0 < \x_i < 1$ is equivalent to
\begin{equation*}
\fp'_i(0) < \alpha \ad_i + \beta w_i < \fp'_i(1),
\end{equation*}
and the only conclusion consistent with~\eqref{eqn:Lemma1} and~\eqref{eqn:Lemma2} is that $\fp'_i(\x_i) =  \alpha \ad_i + \beta w_i$,
or equivalently,
\begin{equation*}
\x_i = {\fp'_i}^{-1}(\alpha \ad_i + \beta w_i).
\end{equation*}

The last possibility corresponds to the case when $\x_i = 1$,
which by equations~\eqref{eqn:Lemma3} and~\eqref{eqn:Lemma2} imply $\fp'_i(1) \leqslant \alpha \ad_i +\beta w_i$.
Next, we check that this latter condition is sufficient for $\x_i=1$.
We first assume $0<\x_i<1$.
In this case, $\lambda_i = \mu_i = 0$ and the dual optimality conditions reduce to $\fp'_i(\x_i) = \alpha \ad_i +\beta w_i$,
which contradicts the fact that $\fp'_i$ is strictly increasing. 
Assuming $\x_i=0$, on the other hand, leads to $\fp'_i(0) \geqslant \alpha \ad_i +\beta w_i$, which runs contrary to the condition $\fp'_i(1) \leqslant \alpha \ad_i +\beta w_i$ and the strict monotonicity of $\fp'_i$.

In summary, $\x_i=0$ if $\alpha \ad_i + \beta w_i \leqslant \fp'_i(0)$, or equivalently, ${\fp'_i}^{-1}(\alpha \ad_i + \beta w_i ) \leqslant 0$;
$\x_i= {\fp'_i}^{-1}(\alpha \ad_i + \beta w_i)$ if $\fp'_i(0) < \alpha \ad_i + \beta w_i <\fp'_i(1)$, or equivalently, $0 <{\fp'_i}^{-1}(\alpha \ad_i + \beta w_i ) < 1$;
and $\x_i=1$ if $\alpha \ad_i + \beta w_i \geqslant \fp'_i(1)$, or equivalently, ${\fp'_i}^{-1}(\alpha \ad_i + \beta w_i ) \geqslant 1$.
Accordingly, it is immediate to obtain the solution form given in the statement.
\EndProof

As mentioned at the beginning of this subsection, the optimization problem presented in the lemma is the same as that of~\eqref{eqn:PrivacyMoneyFunction} but for additively separable, twice differentiable objective functions, with strictly increasing derivatives.
Although these requirements obviously restrict the space of possible privacy functions of our analysis,
the fact is that some of the best known dissimilarity and distance functions satisfy these requirements.
This is the case of
some of the most important examples of Bregman divergences~\cite{Bregman67CMMP}, such as the SED, KL divergence and the Itakura-Saito distance (ISD)~\cite{Itakura68AC}. In the interest of brevity, many of the results shown in this section will be derived only for some of these three particular distance measures.
Due to its mathematical tractability, however, special attention will be given to the SED.

For notational simplicity, hereafter we shall denote by $z_i$ and $\gamma$ the column vectors $(\ad_i, w_i)$ and $(\alpha, \beta)$, respectively.
A compelling result of Lemma~\ref{lem:genericproblem} is the maximin form of the solution and its dependence on the inverse of the derivative of the privacy function.
The particular form that each of the $n$ components of the solution takes, however, hinges
on whether $\ad_i \alpha + w_i \beta$ is greater or less than the value of the derivative of $\fp_i$ at 0 and 1; equivalently, in our vector notation, the lemma shows that the solution is determined by the specific configuration of the $n$ slabs
\begin{equation*}
\nabla \fp(0) \preccurlyeq  z^{\oT} \gamma \preccurlyeq \nabla \fp(1),
\end{equation*}
where $\nabla\fp(0)$ denotes the gradient of $\fp$ at 0,
and $z_i$ are the columns of $z$.
In particular, the $i$-th component of the solution is equal to 0, 1 or ${\fp'_i}^{-1}(z_i^{\oT} \gamma)$ if, and only if, $z_i^{\oT} \gamma\leqslant \fp'_i(0)$, $z_i^{\oT} \gamma \geqslant \fp'_i(1)$, or $\fp'_i(0) < z_i^{\oT} \gamma < \fp'_i(1)$, respectively.

From the lemma, it is clear then that $\gamma$,
which must satisfy the primal equality constraints $\ad^{\oT}\x = 0$ and $w^{\oT}\x = \mu$, is the parameter that configures the point of operation within the $\alpha$-$\beta$ plane where all such halfspaces lie.
Informally, the region of this plane where $\gamma$ falls on is what determines which precise components are 0, 1 and ${\fp'_i}^{-1}(z_i^{\oT} \gamma)$.
Nevertheless, the problem when trying to determine the particular form of each of the $n$ components is
the apparent arbitrariness and lack of regularity of the layout drawn by their corresponding slabs, which makes it difficult to obtain an explicit closed-form solution for any given $\mu, q, p, w$ and $n$. Especially for large values of $n$, conducting a general study of the optimal trade-off between privacy and economic reward becomes intractable.

\begin{figure}[t]
\centering
\includegraphics[scale= 0.98]{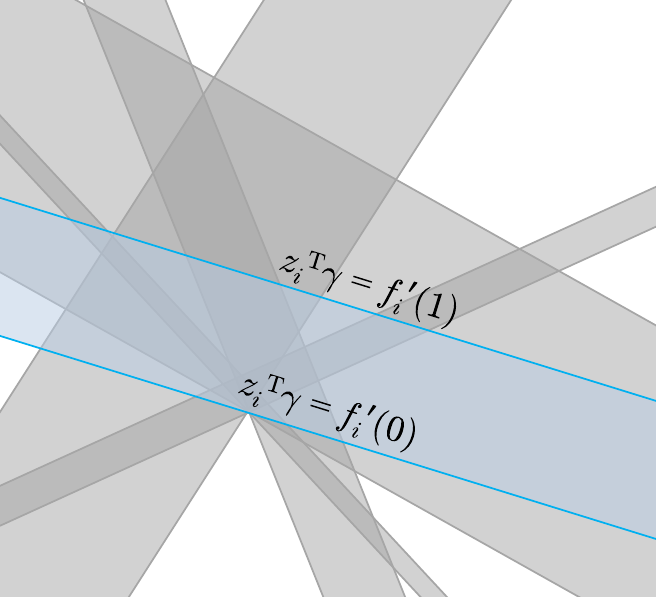}
\caption{Slabs layout on the $\alpha$-$\beta$ plane for $n=6$ categories. Each component of the solution is determined by a slab and, in particular, by the specific $\gamma$ falling on the plane. We show in dark blue the lower and upper hyperplanes of the $i$-th slab.
In general, it will be difficult to proceed towards an explicit closed-form solution and to study the corresponding optimal privacy-money trade-off for any configuration of these slabs and any $\gamma$ and $n$.}
\label{fig:slabs}
\end{figure}

Motivated by all this, our analysis of the solution and the corresponding trade-off focuses on some specific albeit riveting cases of slabs layouts. In particular, Sec.~\ref{sec:Theory:CaseN2} will examine several instantiations of the problem~\eqref{eqn:actualproblem} for small values of $n$. Afterwards, Sec.~\ref{sec:Theory:CaseN2} will tackle the case of large $n$ for some special layouts that will permit us to systematize our theoretical analysis.
Fig.~\ref{fig:slabs} shows a configuration of slabs for $n=6$, and illustrates the conditions that define an optimal strategy.

\subsection{Origin of Lower Hyperplanes}
\label{sec:Theory:Specific}
\noindent
Despite the arbitrariness of the layout depicted by the slabs associated with a particular instantiation of the problem~\eqref{eqn:actualproblem}, next we shall be able to derive an interesting property for some specific privacy functions.
The property in question is related to the need of establishing a fixed point of reference for the geometry of the solutions space.

\begin{proposition}[Intersection of Lower Hyperplanes]
\label{prp:intersection}
In the case when $q \neq p$,
if $\ad_i \fp'_j (0) = \ad_j \fp'_i (0)$ for all $i,j =1,\ldots,n$ and $i\neq j$, then the hyperplanes $z_i^{\oT} \gamma = \fp'_i(0)$ for $i=1,\ldots,n$ all intersect at a single point $O$ on the plane $\alpha$-$\beta$.
\end{proposition}

\BeginProof
Clearly, the consequent of the statement is true if, and only if,
the system of equations $z^{\oT} \gamma = \nabla\fp(0)$ has a unique solution.
We proceed by proving that the rank of the coefficient and augmented matrices is equal to 2 under the conditions stated in the proposition.

On the one hand, recall that $z_i = (\ad_i, w_i)$ is the $i$-th column of $z$, and check that its rank is two if, and only if, $\ad_i w_j \neq \ad_j w_i$ for some $i,j=1,\ldots,n$ and $i\neq j$.
That said, now we show that the consequent of this biconditional statement is true provided that $q\neq p$.
To this end, we assume, by contradiction, that $\sgn(\ad_1)=\cdots=\sgn(\ad_n)$, where $\sgn(\cdot)$ is the sign function. 
If $\ad_i = q_i - p_i > 0$ for $i=1,\ldots,n$, we have $1 = \sum q_i > \sum p_i = 1$, a contradiction.
The case $\ad_i < 0$ for all $i$ leads to an analogous contradiction, and the case $\ad_i=0$ (for all $i$) contradicts the fact that $q \neq p$.
Hence, the condition $q \neq p$ implies that there must exist some indexes $i,j$ with $i\neq j$ such that $\sgn(\ad_i) \neq \sgn(\ad_j)$, which in turn implies that $\ad_i w_j \neq \ad_j w_i$, and that $\rank(z)=2$.

On the other hand, to check the rank of the augmented matrix, observe that the determinant of any 3x3 submatrix with rows $i,j,k$ yields
\begin{align*}
\label{eqn:origincondition}
\deter\left(z|\nabla \fp(0)\right) & =w_i (\ad_j \fp'_k(0) - \ad_k \fp'_j(0))\\
                       & +w_j (\ad_i \fp'_k(0) - \ad_k \fp'_i(0))\\
                       & +w_k (\ad_i \fp'_j(0) - \ad_j \fp'_i(0)).
\end{align*}
From this expression, it is easy to verify that $\rank\left(z|\nabla \fp(0)\right)=2$ if all terms $\ad_i \fp'_j(0) - \ad_j \fp'_i(0)$ with $i\neq j$ vanish, which ensures, by the Rouch\'e-Capelli theorem~\cite{Lang93B}, that there exists a unique solution to~$z^{\oT} \gamma =\nabla\fp(0)$.
\EndProof

The importance of Proposition~\ref{prp:intersection} is obvious: for some privacy functions and distributions $q$ and $p$,
the existence of a sort of origin of coordinates in the slabs layout may reveal certain regularities which may help us systematize the analysis of the solutions space.
For example, a trivial consequence of the intersection of all lower hyperplanes on $O$ is that any $\gamma$ lying on an bounded polyhedron will lead to a solution with at least one component of the form ${\fp'_i}^{-1}(z_i^{\oT} \gamma)$ on its interior. When the assumptions of the above proposition does not satisfy, however, this property may not hold for any $n$ and the choice of the origin may not be evident.

In the next subsections, we shall investigate the optimal trade-off between privacy and money for several particular cases. As we shall see, these cases will leverage certain regularities derived from, or as a result of, said reference point on the $\alpha$-$\beta$ plane. Before that, however, our next result, Corollary~\ref{col:specificorigin}, provides such point for each of the three privacy functions considered in our analysis.
\begin{corollary}
\label{col:specificorigin}
Consider the nontrivial case when $q\neq p$.
The solution to $z^{\oT}\gamma = \nabla \fp(0)$ is unique and yields $(0,0)$ for the squared Euclidean and the Itakura-Saito distances, and $(1,0)$ for the KL divergence.
\end{corollary}

\BeginProof
We obtain the result as a direct application of Proposition~\ref{prp:intersection}.
Note that the gradient of the squared Euclidean and the Itakura-Saito distances vanishes at $\x=0$.
In the case of the KL divergence, $\nabla\fp(0) = (\ad_1,\ldots,\ad_n)$.
Clearly, in the three cases investigated, the condition $\ad_i \fp'_j(0) = \ad_j \fp'_i(0)$ for all $i\neq j$ in the proposition is satisfied,
which implies that the solution is unique.
Then, it is immediate to derive the solutions claimed in the statement.
\EndProof

Although it seems rather obvious, the above corollary actually tells us something of real substance. In particular,
for the three privacy functions under study, $O$ does not depend on a user's profile nor the particular initial distribution chosen.
This result therefore shows the appropriateness of basing our analysis on such functions.

\subsection{Case $n\leqslant 3$}
\label{sec:Theory:CaseN2}
\noindent
We start our analysis of several specific instantiations of the problem~\eqref{eqn:actualproblem} for small values of the number of interest categories $n$. We shall first tackle the case $n=2$ and afterwards the case $n=3$.

The special case $n=2$ reflects a situation in which a user may be willing to group the original set of topics (e.g., business, entertainment, health, religion, sports) into a ``sensitive'' category (e.g., health, religion) and a ``non-sensitive'' category (e.g., business, entertainment, sports), and disclose their interests accordingly. Evidently, this grouping would require that the user specify the same rate $w_i$ for all topics belonging to one of these two categories.
Our next result, Theorem~\ref{thr:casen2}, presents a closed-form solution to the minimization problem involved in the definition of function~\eqref{eqn:PrivacyMoneyFunction} for this special case. As we shall see now, this result can be derived directly from the primal feasibility conditions.

\begin{theorem}[Case $n=2$, and SED and KL divergence]
\label{thr:casen2}
Let $\fp:[0,1]\times [0,1] \to\mathbb{R}_+$ be continuous on the interior of its domain.
\begin{EnumerateRoman}
\item For any $\mu \in [0,\mumax]$ and $i=1,2$, the optimal disclosure strategy is $\x^*_i = \frac{\mu}{\mumax}$.
\item In the case of the SED and KL divergence, the corresponding, minimum distance yields the privacy-money functions
\begin{align*}
        &\cR_{\textnormal{SED}}(\mu) = 2\left(\ad_i \frac{\mu}{\mumax}\right)^2 \textnormal{ and }\\
        &\cR_{\textnormal{KL}}(\mu) = \sum_{i=1}^2 \left(\ad_i \frac{\mu}{\mumax} + p_i\right) \log\left(\frac{\ad_i\, \mu/\mumax}{p_i} + 1\right).
\end{align*}
\end{EnumerateRoman}
\end{theorem}
\BeginProof
Since $n=2$, we have that $\ad_1 = - \ad_2$, which, by virtue of the primal condition $\sum \ad_i \x^*_i = 0$, implies that $\x^*_1 = \x^*_2$.
Then, from the other primal condition $\sum w_i \x^*_i = \mu$, it is immediate to obtain the solution claimed in assertion~(i) of the theorem.
Finally, it suffices to substitute the expression of $\x^*$ into the functions $\fp_{\textnormal{SED}}(\x_i) = \sum_i (t^*_i - p_i)^2$ and
$\fp_{\textnormal{KL}}(t^*,p) = \sum_i t^*_i \log {t^*_i}/p_i$, to derive the optimal trade-off function $\cR(\mu)$ in each case.
\EndProof

In light of Theorem~\ref{thr:casen2},
we would like to remark the simple, linear form of the solution, which, more importantly, is valid for a set of privacy functions which is larger than that considered in Lemma~\ref{lem:genericproblem}. In particular, not only the KL divergence, the squared Euclidean and the Itakura-Saito distances satisfy the conditions of this theorem, but also many others which are not differentiable (e.g., total variation distance) nor additively separable (e.g., Mahalanobis distance).

Another straightforward consequence of Theorem~\ref{thr:casen2} is that the optimal strategy implies revealing both categories (e.g., sensitive and non-sensitive) simultaneously and with the same level of disclosure. In other words, if a user decides to show a fraction of their interest in one category, that same fraction must be disclosed on the other category so as to attain the maximum level of privacy protection.

Before proceeding with Theorem~\ref{thr:casen3},
first we shall introduce what we term \emph{money thresholds},
two rates that will play an important role in the characterization of the solution to the minimization problem~\eqref{eqn:actualproblem} for $n=3$.
Also, we shall introduce some definitions that will facilitate the exposition of the aforementioned theorem.

For $i=1,\ldots,n$, denote by $m_i$ the slope of vector $z_i$, i.e., $m_i=\frac{w_i}{\ad_i}$.
Let $\overline{m}_i$ and $\sigma^2_{m_i}$ be the arithmetic mean and variance of all but the $i$-th slope.
When the subindex $i\not\in\sX$, observe that the mean and variance are computed from all slopes.
Accordingly, define the \emph{money thresholds} $\mu_j$ as
\begin{equation*}
\mu_j =  \min_{i\neq 2j} \frac{(j+1)\,\ad_i\,\sigma^2_{m_{2j}}}{m_i - \overline{m}_{2j}}
\end{equation*}
for $j=1,2$.

Additionally, we define the \emph{relative coefficient of variation} of the ratio $w_i/\ad_i$ as
\vspace{-0.02cm}
\begin{equation}
v_{i,j} = \frac{m_i - \overline{m}_j}{\sigma^2_{m_{j}}}
\end{equation}
\vspace{-0.02cm}
for $i,j=1,\ldots,n$, which may be regarded as the inverse of the index of dispersion~\cite{Shao99B}, a measure commonly utilized in statistics and probability theory to quantify the dispersion of a probability distribution.
As we shall show in the following result, our coefficient of variation will determine the closed-form expression of the optimal disclosure strategy.
\noindent
\begin{theorem}[Case $n=3$ and SED]\label{thr:casen3}
For $n=3$ and the SED function, assume without loss of generality $m_1 \geqslant m_2 \geqslant m_3$.
Either $w_{j+1} \leqslant \ad_{j+1}\,\overline{m}_{j+1}$ for $j=1$ and $m_1 > m_3$, or $w_j > \ad_j\,\overline{m}_j$ for $j=2$.
For the corresponding index~$j$ and for any $\mu \leqslant \mu_j$, the optimal disclosure strategy is
$$\x^*_i=\left\{ \begin{array}{l@{,\quad}l}
                                           \frac{v_{i,2j}}{(j+1)\ad_i}\,\mu  \,\, & i \neq 2j \\
                                            0 \,\, & i=2j
               \end{array}\right.,$$
and the corresponding, minimum SED yields the privacy-money function
\begin{equation*}
\cR_{\textnormal{SED}}(\mu) = \frac{\mu^2}{(j+1)\,\sigma_{m_{2j}}^2}.
\end{equation*}
\end{theorem}

\BeginProof
It is straightforward to verify that the SED function exposes the structure of the optimization problem addressed in Lemma~\ref{lem:genericproblem}.
Note that, according to the lemma, the components of the solution such that $0<\x_i<1$ for some $i=1,2,3$ are given by the inverse of the privacy function and yield
$${\fp'_i}^{-1}(\alpha \ad_i + \beta w_i) = \frac{\alpha}{2\,\ad_i} + \frac{w_i\,\beta}{2\,\ad_i^2}.$$
To check that a solution does not admit only one positive component,
simply observe that the system of equations composed of the two primal equality conditions $\sum_i \ad_i\,\x_i = 0$ and $\sum_i w_i \x_i =\mu$ is inconsistent.

Having shown that there must be at least two positive components, we apply such primal equality conditions to a solution with $0<\x_1,\x_3<1$.
To verify these two equalities are met, first note that the former
is equivalent to $\alpha + \beta\,\overline{m}_2 = 0$,
and the latter can be written equivalently as
\vspace{-0.02cm}
\begin{equation*}
\alpha\,\overline{m}_2 + \frac{\beta}{2}\,\sum_{i=1,3} m_i^2 = \mu.
\end{equation*}
\vspace{-0.02cm}
Then, observe that the condition $m_1 > m_3$ in the theorem ensures that the determinant of the homogeneous system is nonzero,
and, accordingly, that the Lagrange multipliers that solve these two equations are
\vspace{-0.02cm}
\begin{equation}
\label{eqn:alphabeta}
\alpha = - \frac{\overline{m}_2}{\sigma_{{m}_2}^2}\mu\,\,\textnormal{ and }\,\,\beta= \frac{1}{\sigma_{{m}_2}^2}\,\mu.
\end{equation}
\vspace{-0.02cm}
Finally, it suffices to substitute the expressions of $\alpha$ and $\beta$ into the function ${\fp'_i}^{-1}$,
to obtain the solution with two nonzero optimal components claimed in the theorem.

Next, we derive the conditions under which this solution is defined. With this aim, just note that the inequalities $z_1^{\oT} \gamma > \fp'_1(0)$ and $z_3^{\oT} \gamma > \fp'_3(0)$ are equivalent to $\ad_1\left(m_1 - \overline{m}_2\right) > 0$ and $\ad_3\left(m_3 - \overline{m}_2\right) > 0$, respectively.
On the other hand, $\x_2 = 0$ if, and only if, $z_2^{\oT} \gamma \leqslant \fp'_2(0)$, or equivalently, $\ad_2 \left(m_2 - \overline{m}_2\right) \leqslant 0$.

We now show that when there are two components $0<\x_i,\x_j<1$, then $i=1$ and $j=3$. To this end, we shall examine the case $0<\x_2,\x_3<1$ and $\x_1=0$. The other possible case, $0<\x_1,\x_2<1$ and $\x_3=0$, proceeds along the same lines and is omitted.

First, though, we shall verify that $\ad_1 \geqslant 0$, a condition that will be used later on. We proceed by contradiction. Since $w_i > 0$ for all $i$, a negative $\ad_1$ implies, by the ordering assumption $m_1 \geqslant m_2 \geqslant m_3$, that $\ad_2,\ad_3 < 0$. But having $\ad_i<0$ for $i=1,2,3$ leads us to the contradiction $0 > \sum_i \ad_i = \sum_i q_i - \sum_i p_i =0$. Consequently, $\ad_1$ is nonnegative, but by virtue of~\eqref{eqn:NoEqual}, it follows that $\ad_1 > 0$.

Having verified the positiveness of $\ad_1$, next we contemplate the case when $0<\x_2,\x_3<1$ and $\x_1 = 0$. Note that, in this case, the condition $\x_1 =0$ holds if, and only if, $\ad_1 \left(m_1 - \overline{m}_1 \right) \leqslant 0$.
However, since $\ad_1 >0$, we have that $m_1 \leqslant \frac{1}{2}\left(m_2 + m_3\right)$,
which contradicts the fact that $m_1 \geqslant m_2 \geqslant m_3$ and $m_1>m_3$.
Consequently, it is not possible to have $0<\x_2,\x_3<1$ and $\x_1=0$.
The case when $0<\x_1,\x_2<1$ and $\x_3=0$ leads to another contradiction and the conclusion that
$0<\x_1,\x_3<1$ and $\x_2=0$.

Next, we check the validity of the conditions under which this solution is defined. Recall that these conditions are $\ad_1\left(m_1 - \overline{m}_2\right) > 0$, $\ad_3\left(m_3 - \overline{m}_2\right) > 0$ and $\ad_2\left(m_2 - \overline{m}_2\right) \leqslant 0$.
It is easy to verify that the former two inequalities hold, since the arithmetic mean is strictly smaller (greater) than the extreme value $m_1$ ($m_3$); the strictness of the inequality is due to the assumption $m_1 > m_3$ in the statement.
On the other hand, the latter inequality is the condition assumed in the statement of the theorem.
Therefore, we have $0<\x_1,\x_3<1$ and $\x_2=0$ if, and only if, $w_2 \leqslant \ad_2\, \overline{m}_2$.

Next, we turn to the case when $0<\x_1,\x_2,\x_3<1$. By applying the two primal equality constraints of the optimization problem~\eqref{eqn:actualproblem}, we obtain the system of equations
\begin{equation*}
\frac{3}{2}
\begin{bmatrix}
1                          & \overline{m}_0 \\
\overline{m}_0 & \frac{1}{3}\sum_{i=1}^3 m_i^2
\end{bmatrix}\,
\begin{bmatrix}
\alpha\\
\beta
\end{bmatrix} =
\begin{bmatrix}
0\\
\mu
\end{bmatrix},
 \end{equation*}
and note that the solution is unique on account of the fact that $\sgn(\ad_i) \neq \sgn(\ad_j)$ for some $i,j=1,2,3$ and $i\neq j$, which implies that $\sigma_{{m}_0}^2>0$.
Substituting the values
\begin{equation}
\label{eqn:alphabeta2}
\alpha = - \frac{2\,\overline{m}_0}{3\,\sigma_{{m}_0}^2}\mu\,\,\textnormal{ and }\,\,\beta= \frac{2}{3\,\sigma_{{m}_0}^2}\,\mu
\end{equation}
into ${\fp'_i}^{-1}(z_i^{\oT} \gamma)$ gives the expression of the optimal disclosure strategy stated in the theorem for $0<\x_1,\x_2,\x_3<1$.

Now, we examine the necessary and sufficient conditions for this optimal strategy to be possible, which, according to the lemma, are $0< z_i^{\oT} \gamma < 2\,\ad_i$ for $i=1,2,3$.  To this end, note that the left-hand inequalities can be recast as
$\ad_i\,(m_i - \overline{m}_0) > 0$,
for $i=1,2,3$.
We immediately check that the inequalities for $i=1$ and $i=3$ hold, as the mean is again strictly smaller (greater) than the extreme value $m_1$ ($m_3$). The strictness of these two inequalities is due to the fact that $\sum_{i=1}^3 \ad_i = 0$ and the assumption~\eqref{eqn:NoEqual}.
On the other hand, observe that
$$\sgn(m_2 - \overline{m}_0) = \sgn(m_2 - \overline{m}_2),$$
and therefore that the condition $\ad_2\,\left(m_2 - \overline{m}_0\right) >0$ is equivalent to $\ad_2\,\left(m_2 - \overline{m}_2\right) >0$.
That said, note that $\ad_2\,\left(m_2 - \overline{m}_2\right) >0$ is the negation of the condition for having a solution with two nonzero components smaller than one.
Accordingly, we have either two or three components of this form, as stated in the theorem.

To show the validity of the solution in terms of $\mu$, observe that, for $w_2 \leqslant \ad_2\,\overline{m}_2$, the parameterized line $\left(\alpha(\mu),\beta(\mu)\right)$ moves within the space determined by the intersection of the slabs 1 and 3. To obtain the range of validity of a solution such that $0<\x_1,\x_3<1$ and $\x_2=0$, we need to find the closest point of intersection (to the origin) with either the upper hyperplane 1 or the upper hyperplane 3.
Put differently, we require finding the minimum $\mu$ such that either $z_1^{\oT} \gamma = \fp'_1(1)$ or $z_3^{\oT} \gamma = \fp'_3(1)$.
By plugging the values of $\alpha$ and $\beta$ given in~\eqref{eqn:alphabeta} into these two equalities, it is straightforward to derive the money threshold~$\mu_1$.
We proceed similarly to show the interval of validity~$[0,\mu_2]$ in the case when $w_2 >\ad_2\,\overline{m}_2$, bearing in mind that now $\alpha$ and $\beta$ are given by~\eqref{eqn:alphabeta2}.

To conclude the proof, it remains only to write $\cR(\mu)$ in terms of the optimal apparent distribution, that is,
$\cR(\mu)=\sum_{i=1}^n \left(t_i - p_i\right)^2 = \sum_{i=1}^n \ad_i^2\,\x_i^2$,
and from this, it is routine to obtain the expression given at the end of the statement.
\EndProof

Theorem~\ref{thr:casen3} provides an explicit closed-form solution to the problem of optimal profile disclosure, and characterizes the corresponding trade-off between privacy and money.
Although it rests on the assumption that $\mu<\mu_1,\mu_2$ and ---for the sake of tractability and brevity--- tackles only the case of SED,
the provided results shed light on the understanding of the behavior of the solution and the trade-off, and enables us to establish interesting connections with concepts from statistics and estimation theory.

In particular, the most significant conclusion that follows from the theorem
is the intuitive principle upon which the optimal disclosure strategy operates.
On the one hand, in line with the results obtained in Theorem~\ref{thr:casen2}, the solution does not admit only one positive component: we must have either two or three active components.
On the other hand, and more importantly, the optimal strategy is linear with the relative coefficient of variation of the ratio $w_i/\ad_i$, a quantity that is closely related to the index of dispersion, also known as Fano factor\footnote{The difference with respect to these quantities is that our measure of dispersion inverses the ratio variance to mean, and also reflects the deviation with the particular value attained by a given component.}.

The solution, however, does not only depend on $v_{i,j}$ but also on the difference between the interest value of the actual profile and that of the initial PMF. Essentially, the optimized disclosure works as follows.
We consider the category $i$ with the largest value $w_i$, which in practice may correspond to the most sensitive category.
For that category,
if $\ad_i$ is small and $m_i$ is the ratio that deviates the most from the mean value ---relative to the variance---, then the optimal strategy suggests disclosing the profile mostly in that given category. This conforms to intuition since, informally, revealing small differences $q_i-p_i$ when $w_i$ is large may be sufficient to satisfy the broker's demand, i.e., the condition $\sum_i w_i \x_i =\mu$, and this revelation may not have a significant impact on user privacy\footnote{Bear in mind that, when using $\fp_{\textnormal{SED}}$ to assess privacy, small values of $\ad_i$ lead to quadratically small values of privacy risk.}.
On the other hand, if $\ad_i$ is comparable to $w_i$, and $m_i$ is close to the mean value, then $\x^*$ recommends that the user give priority to other categories when unfolding their profile.

Also, from this theorem we deduce that the optimal trade-off depends quadratically on the offered money, exactly as with the case $n=2$, and inversely on the variance of the ratios $m_1, m_2, m_3$.

Last but not least,
we would like to remark that, although Theorem~\ref{thr:casen3} does not completely\footnote{That is, for all values of $\mu$.} characterize the optimal strategy nor the corresponding trade-off for any $q$, $p$, $w$ and $\mu$ for $n=3$,
the proof of this result does show how to systematize the analysis of the solution for any instance of those variables. Sec.~\ref{sec:Examples} provides an example that illustrates this point.

\subsection{Case $n\geqslant 3$ and Conical Regular Configurations}
\label{sec:Theory:CaseStar}
\noindent
In this subsection, we analyze the privacy-money trade-off for large values of $n$, starting from 3. 
To systematize this analysis, however, we shall restrict it to a particular configuration of the slabs layout, defined next. Then, Proposition~\ref{prp:CaseStar} will show an interesting property of this configuration, which will allow us to derive an explicit closed-form expression of both the solution and trade-off for an arbitrarily large number of categories.
\begin{definition}
\label{def:CaseStar}
For a given $q,\,p,\,w$ and $n\geqslant3$, let $\mathscr{C}$ be the collection of slabs on the plane $\alpha$-$\beta$ that determines the corresponding solution to~\eqref{eqn:actualproblem} stated in Lemma~\ref{lem:genericproblem}. Without loss of generality, assume $\frac{1}{m_1} > \cdots > \frac{1}{m_n}$.
Define $A_i$, $b_i$ and $b'_i$ as
\begin{equation*}
A_i =
\begin{bmatrix}
z_i^{\oT}\\
z_{i-1}^{\oT}\\
z_1^{\oT}
\end{bmatrix},
b_i =
\begin{bmatrix}
 \fp'_i(0)\\
\fp'_{i-1}(1)\\
\fp'_1(1)
\end{bmatrix}
\textnormal{and}\,\,
b'_i =
\begin{bmatrix}
\fp'_i(1)\\
\fp'_{i-1}(1)\\
\fp'_1(0)
\end{bmatrix}.
 \end{equation*}
Then, $\mathscr{C}$ is called a \emph{conical regular configuration} if each of the system of equations $A_i\,\gamma = b_i$ and $A_i\,\gamma = b'_i$ for $i=3,\ldots,n$ has a unique solution.
\end{definition}

\begin{figure}[t]
\centering
\includegraphics[scale= 0.67]{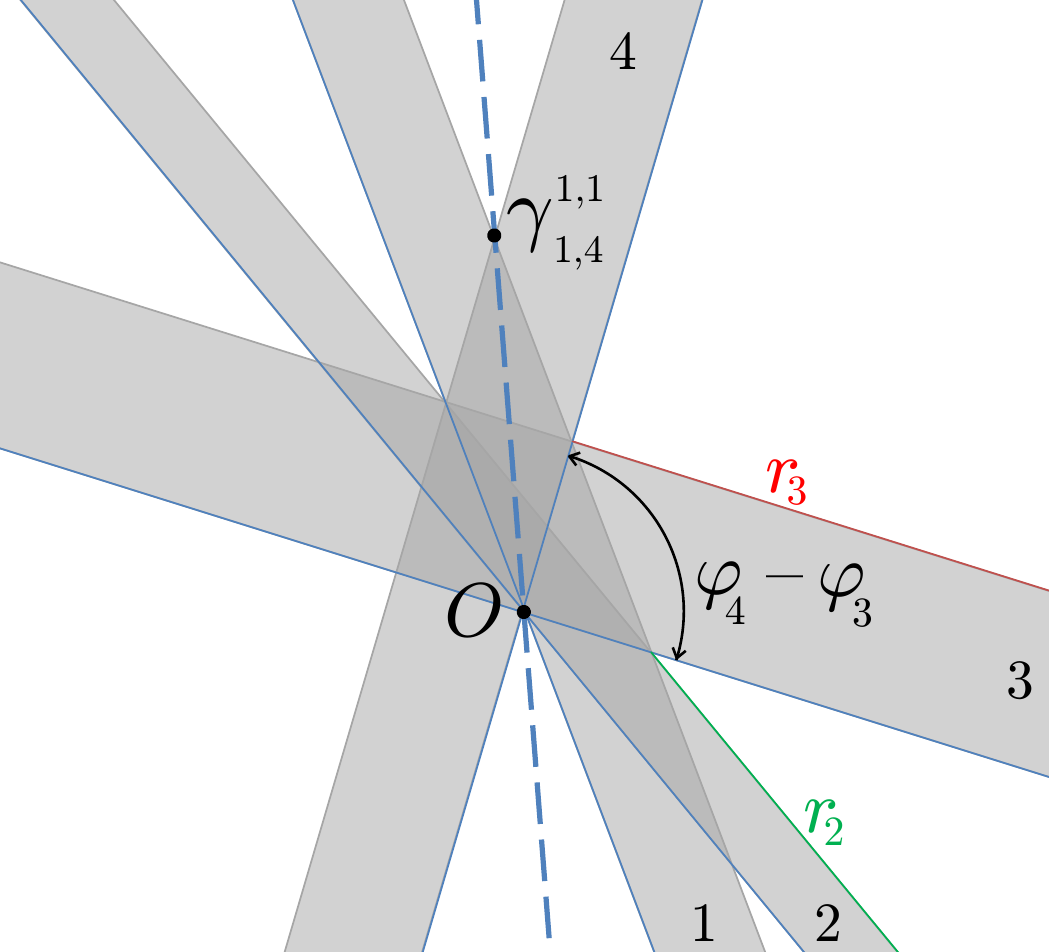}
\caption{A conical regular configuration for $n=4$ on the $\alpha$-$\beta$ plane.
In this figure, we show the segments of hyperplanes $r_2(\varphi)$ and $r_3(\varphi)$, given respectively by the angular coordinates $\varphi_2\leqslant \varphi \leqslant \varphi_3$ and $\varphi_3\leqslant \varphi \leqslant \varphi_4$.
The cone defined by $r\geqslant0$ and $\varphi_3\leqslant \varphi \leqslant \varphi_4$ is intersected by the upper hyperplanes 1, 2 and 3. However, neither of these hyperplanes intersect among themselves on the interior of the cone in question.}
\label{fig:slabsproperties}
\end{figure}

\begin{proposition}
\label{prp:CaseStar}
Suppose that there exists a conical regular configuration $\mathscr{C}$ for some $q,\,p,\,w$ and $n$.
Denote by $\gamma_{i,j}^{a,b}$ the unique solution to
\begin{equation*}
\begin{cases}
        z_i^{\oT}\,\gamma = \fp'_i(a)\\
        z_j^{\oT}\,\gamma = \fp'_j(b)
\end{cases}
\end{equation*}
for $i,j=1,\ldots,n$ with $i\neq j$, and $a,b\in\{0,1\}$.
Assume $\fp'_i(0)\neq \fp'_i(1)$ for all $i$.
Then, except for $\gamma_{1,n}^{1,1}$, $\mathscr{C}$ satisfies
\begin{equation}
\label{eqn:internalprpstar}
z_k^{\oT}\,\gamma_{i,j}^{a,b} = \fp'_k(0)
\end{equation}
for some $k=1,\ldots,n$ and all $i\neq j$.
\end{proposition}

\BeginProof
The existence and uniqueness of $\gamma_{i,j}^{a,b}$ is guaranteed by the fact that $\frac{1}{m_1} > \cdots > \frac{1}{m_n}$.
The property stated in the proposition follows from the fact that the systems of equations $A_i\,\gamma = b_i$ and $A_i\,\gamma = b'_i$ for $i=3,\ldots,n$ have a unique solution.

The systems of equations of the form $A_i\,\gamma = b_i$ ensure that $\gamma_{i,1}^{1,1} = \gamma_{i+1,1}^{0,1}$ for $i=2,\ldots,n-1$.
Obviously, any $\gamma_{i,j}^{a,b}$ such that $a=0$ or $b=0$ with $i\neq j$ satisfies~\eqref{eqn:internalprpstar} for $k=i$ or $k=j$.
Accordingly, we just need to prove the case $a=b=1$. 

Suppose $i>j$. Note that $A_i\,\gamma = b'_i$ implies, on the one hand, that
$$\gamma_{i,i-1}^{1,1} = \gamma_{i-1,1}^{1,0} = \gamma_{i-2,1}^{1,0} =\cdots = \gamma_{j,1}^{1,0},$$
and on the other hand, that $\gamma_{j,j-1}^{1,1} = \gamma_{j,1}^{1,0}$.
Thus, $\gamma_{i,i-1}^{1,1} = \gamma_{j,j-1}^{1,1}$, from which it follows that $\gamma_{i,j}^{1,1} = \gamma_{j,1}^{1,0}$.
The exception, i.e., $z_k^{\oT}\,\gamma_{1,n}^{1,1} \neq \fp'_k(0)$ for all $k=1,\ldots,n$, is justified by the conditions $\fp'_i(0)\neq \fp'_i(1)$ for all $i$, which guarantee that all slabs have nonempty interiors, and the strict ordering $\frac{1}{m_1} > \cdots > \frac{1}{m_n}$.
\EndProof

The previous proposition shows a remarkable feature of the conical regular configuration:
at a practical level, the fact that all intersections on the plane $\alpha$-$\beta$ (except $\gamma_{1,n}^{1,1}$) lie on lower hyperplanes suggests utilizing these hyperplanes, parameterized in polar coordinates with respect to the origin $O$, to efficiently delimit the solutions space. 
In other words, in our endeavor to systematize the study of the solution and trade-off, it may suffice to use a reduced number of cases, bounded by angles and segments of hyperplanes.

On the other hand and from a geometric standpoint, any consecutive pair of lower hyperplanes defines a cone without intersections in its interior; hence the name of the configuration.
Finally, as the slabs are sorted in increasing order of their slopes, we can go counter-clockwise from slab 1 to $n$, and start again at the line through $O$ and $\gamma_{1,n}^{1,1}$, which serves as a reference axis.

Before we continue examining this concrete configuration, we shall introduce some notation.
Let $\varphi$ and $r$ be the polar coordinates of $\gamma$.
Define the angle thresholds $\varphi_k$ as
\begin{equation*}
\varphi_k=\left\{ \begin{array}{l@{,\,\,\,}l}
                                            \arctan -\ad_k/w_k & k=1,\ldots,n\\
                                            \arctan \frac{\ad_1\,\fp'_n(1) - \ad_n\,\fp'_1(1)}{w_n\,\fp'_1(1) - w_1\,\fp'_n(1)}& k=n+1 \\
                                            \varphi_{k-n-1} + \pi & k=n+2,\ldots, 2n+1
                \end{array}\right.,
\end{equation*}
and the segments of upper hyperplanes $r_j$ as
\begin{equation*}
r_j(\varphi) = \frac{\fp'_j(1)}{z_j^{\oT}\begin{bmatrix}
\cos \varphi \\
\sin \varphi
\end{bmatrix}}
\end{equation*}
for $j=1,\ldots,n$.
Note that $\varphi_{n+1}$ is the angular coordinate of $\gamma_{1,n}^{1,1}$.
Occasionally, we shall omit the dependence of these line segments on the angular coordinate~$\varphi$.
Figure~\ref{fig:slabsproperties} illustrates these coordinates and segments on a conical regular configuration for $n=4$.

Our next result, Lemma~\ref{lem:star}, provides a parametric solution in the special case when the slabs layout exhibits such configuration. The solution is determined by the aforementioned thresholds and line segments, and is valid for any privacy function satisfying the properties stated in Lemma~\ref{lem:genericproblem}. As we shall show next, this result will be instrumental in proving Theorem~\ref{thr:star}.
\begin{lemma}[Conical Regular Configurations]\label{lem:star}
Under the conditions of Lemma~\ref{lem:genericproblem}, assume that there exists a conical regular configuration.
Consider the following cases:
\begin{itemize}
\item [(a)] $\varphi_k <\varphi \leqslant \varphi_{k+1}$ for $k=1$ and,
                             either $r < r_j$ for $j=1$
                             or $r_{j-1} \leqslant r$ for $j=2$;
and $\varphi_k <\varphi \leqslant \varphi_{k+1}$ for $k=2$ and,
                              either $r < r_j$ for $j=1$,
                              or $r_{j-1} \leqslant r < r_j$ for $j=2$,
                              or $r\geqslant r_{j-1}$ for $j=3$.

\item [(b)] $\varphi_k <\varphi \leqslant \varphi_{k+1}$ for some $k=3,\ldots,n$ and,
                                                      either $r <r_{j+1}$ for $j=1$,
                                                      or $r_j \leqslant r < r_{j+1}$ for some $j=2,\ldots,k-2$,
                                                      or $r_j \leqslant r < r_{j+2\Mod{k}}$ for $j=k-1$,
                                                      or $r_{j+1\Mod{k}} \leqslant r < r_j$ for $j=k$,
                                                      or $r\geqslant r_{j-1}$ for $j=k+1$.

\item [(c)] $\varphi_k <\varphi < \varphi_{k+1}$ for $k=n+1$ and,
                                                      either $r <r_{j+1}$ for $j=1$,
                                                      or $r_j \leqslant r < r_{j+1}$ for some $j=2,\ldots,n-1$,
                                                      or $r_j \leqslant r < r_1$ for $j=n$,
                                                     or $r\geqslant r_{j-n}$ for $j=n+1$.

\item [(d)] $\varphi_k \leqslant \varphi < \varphi_{k+1}$ for some $k=n+2,\ldots,2n$ and,
                                                      either $r <r_{n-j+1}$ for $j=1$,
                                                      or $r_{n-j+2} \leqslant r < r_{n-j+1}$ for some $j=2,\ldots, 2n-k+1$, 
                                                      or $r\geqslant r_{n-j+2}$ for $j=2(n+1) - k$.
\end{itemize}
\noindent
Let $\x^*$ be the solution to the problem~\eqref{eqn:actualproblem}. Accordingly,
\begin{itemize}
\item[(i)] in cases (a) and (b), and for the corresponding indexes $k$ and $j$,
\begin{equation*}
\hspace*{-0.5cm}
\x^*_i=\left\{\begin{array}{l@{}l}
                                           0, &\,\,\, i = k+1,\ldots,n\\
                                           {\fp'_i}^{-1}\left(z_i^{\oT}\gamma\right),
                                           &\begin{array}{l@{}l}i=1 \textnormal{ and } i=j+1,\ldots,k &\,\,\textnormal{if}\,j<k\\
                                                                                  i=j,\ldots,k &\,\,\textnormal{if}\,j= k\end{array};\\
                                           1, &\begin{array}{l@{}l}i=2,\ldots,j &\,\,\textnormal{if}\,j<k\\
                                                                                   i=1,\ldots,j-1 &\,\,\textnormal{if}\,j\geqslant k\end{array}
               \end{array}\right.
\end{equation*}
\item[(ii)] in case (c), and for the corresponding indexes $k$ and $j$, the solution is obtained by exchanging the indexes $i=1$ and $i=n$ of the solution given for case (b) and $k=n$;
\item[(iii)] in case (d), and for the corresponding indexes $k$ and~$j$,
$$\x^*_i=\left\{\begin{array}{l@{,\quad}l}
                                           0, &i =1, \ldots,k-n-1\\
                                           {\fp'_i}^{-1}\left(z_i^{\oT}\gamma\right) &i=k-n,\ldots,n-j+1\\
                                           1, &i=n-j+2,\ldots,n
               \end{array}\right..$$
\end{itemize}
\end{lemma}

\BeginProof
From Proposition~\ref{prp:CaseStar}, we have that the conditions $r\geqslant0$ and $\varphi_k \leqslant \varphi \leqslant \varphi_{k+1}$ for any single $k=1,\ldots,n-1,n+2,\ldots,2n$ yield a cone where no intersection of hyperplanes occurs in its interior. Clearly, we also note that each cone is bounded by two consecutive lower hyperplanes and intersected only by upper hyperplanes.
It is easy to verify that the number of intersecting upper hyperplanes is $k$ and $2n-k+1$ for $k=1,\ldots,n$ and $k=n+2,\ldots,2n$, respectively.

That said, all cases in the lemma are an immediate consequence of Lemma~\ref{lem:genericproblem}.
We only show statement~(iii).
With this aim, observe that, for any $k=n+2,\ldots,2n$, the condition $\varphi_k \leqslant \varphi < \varphi_{k+1}$ is equivalent
to $\varphi_{k-n-1} \leqslant \varphi + \pi < \varphi_{k-n}$,
which means that, for a given $k$, the corresponding cone is bounded by the lower hyperplanes $k-n-1$ and $k-n$
and thus
\begin{gather*}
z_{k-n-1}^{\oT}\begin{bmatrix}
r\cos\varphi \\
r\sin\varphi
\end{bmatrix} \leqslant \fp'_{k-n-1}(0).
\end{gather*}
Since a conical configuration satisfies $\frac{1}{m_1} > \cdots > \frac{1}{m_n}$,
then
\begin{gather}
\label{lemma:star:intermediate}
{\fp'_1}^{-1}\left(z_1^{\oT}\gamma\right),\ldots,\fp'^{-1}_{k-n-1}\left(z_{k-n-1}^{\oT}\gamma\right)\leqslant 0,
\end{gather}
and accordingly $\x_1=\cdots=\x_{k-n-1}=0$.

On the other hand, for a given $\varphi \in [\varphi_k,\varphi_{k+1}]$, note that the parameterized line $(r\cos \varphi,r\sin \varphi)$ intersects the sequence of line segments $r_n, r_{n-1},\ldots,r_{k-n}$ when $r$ goes from 0 to $\infty$. This shows the order of the line segments specified in case~(d).

Having checked this, note that when $r_{n-j+2} \leqslant r < r_{n-j+1}$ for some $j=2,\ldots, 2n-k+1$, we have
\begin{gather*}
\fp'^{-1}_{n-j+2}\left(z_{n-j+2}^{\oT}\gamma\right),\ldots, {\fp'_n}^{-1}\left(z_n^{\oT}\gamma\right)\geqslant 1,
\end{gather*}
and thus
$\x_{n-j+2}=\cdots = \x_n=1$.
From~\eqref{lemma:star:intermediate}, it follows that $\x_i = 0$ for $i=1,\ldots,k-n-1$, and then that the rest of the components $i=k-n,\ldots, n-j+1$ must be of the form $\fp'^{-1}_{i}\left(z_{i}^{\oT}\gamma\right)$.
\EndProof

Our previous result, Lemma~\ref{lem:star}, shows that the specific arrangement of the lower and upper hyperplanes of a conical regular configuration makes polar coordinates particularly convenient for analyzing the solution to the optimization problem at hand.
The lemma takes advantage of the regular structure of such configuration, and is used in Theorem~\ref{thr:star} as a stepping stone to derive an explicit closed-form solution for $n\geqslant3$.
To be able to state our next result concisely, we introduce some auxiliary definitions.

Denote by $\cA_i=\sum_{k=i}^n \ad_k$ and $\cW_i=\sum_{k=i}^n w_k$ the complementary cumulative functions of $\ad$ and~$w$.
For $k=n+2,\ldots, 2n$ and $j=1,\ldots,2(n+1)-k$, define the set $\cS(k,j)=\{1,\ldots,k-n-1,n-j+2,\ldots,n\}$. In line with the definition given for case $n\leqslant3$ in Sec.~\ref{sec:Theory:CaseN2}, denote by $\overline{m}_{S(k,j)}$ and $\sigma^2_{m_{S(k,j)}}$ the arithmetic mean and variance of the sequence $(m_i)_{i\in \sX\setminus\cS(k,j)}$.
Similarly to Sec.~\ref{sec:Theory:CaseN2}, we define a sequence of \emph{money thresholds}
\begin{equation*}\label{eqn:thresholds:star}
\mu_{k,j} = \cW_{n-j+2} - \cA_{n-j+2}\Big(\overline{m}_{S(k,j)} + \frac{\sigma^2_{m_{S(k,j)}}}{\overline{m}_{S(k,j)} - m_{k-n-1}}\Big),
\end{equation*}
for $k=n+2,\ldots, 2n$ and $j=1,\ldots,2(n+1)-k$.

\begin{theorem}\label{thr:star}
Assume that there exists a conical regular configuration for some $q,\,p,\,w$ and $n$.
For any $k=n+2,\ldots, 2n$ and $j=1,\ldots,2(n+1)-k$ such that $\mu_{k+1,j}<\mu_{k,j}$,
and for any $\mu\in(\mu_{k+1,j},\mu_{k,j}]$,
the optimal disclosure strategy for the SED function is $\x^*_i= 0$ for $i=1,\ldots,k-n-1$,
\begin{multline*}
\x^*_i= \frac{1}{\ad_i\,\left(n-\cardS\right)}\Big(v_{i,S(k,j)} \left(\mu - \cW_{n-j+2} \right.\\
\left.+ \cA_{n-j+2}\, \overline{m}_{S(k,j)}\right) - \cA_{n-j+2}\Big)
\end{multline*}
for $i=k-n,\ldots,n-j+1$,
and $\x^*_i= 1$ for $n-j+2,\ldots,n$.
\end{theorem}

\BeginProof
The proof parallels that of Theorem~\ref{thr:casen3} and we sketch the essential points.

\begin{figure*}[htb]%
\centering\hspace*{\fill}
\subfigure[$\mu= \$0$, $\cR(\mu)=0$, $\cR(\mu)/\cR(\mumax)=0$, $\delta^*=(0,0,0)$, $t^*=p$.]%
{\includegraphics[scale=0.37]{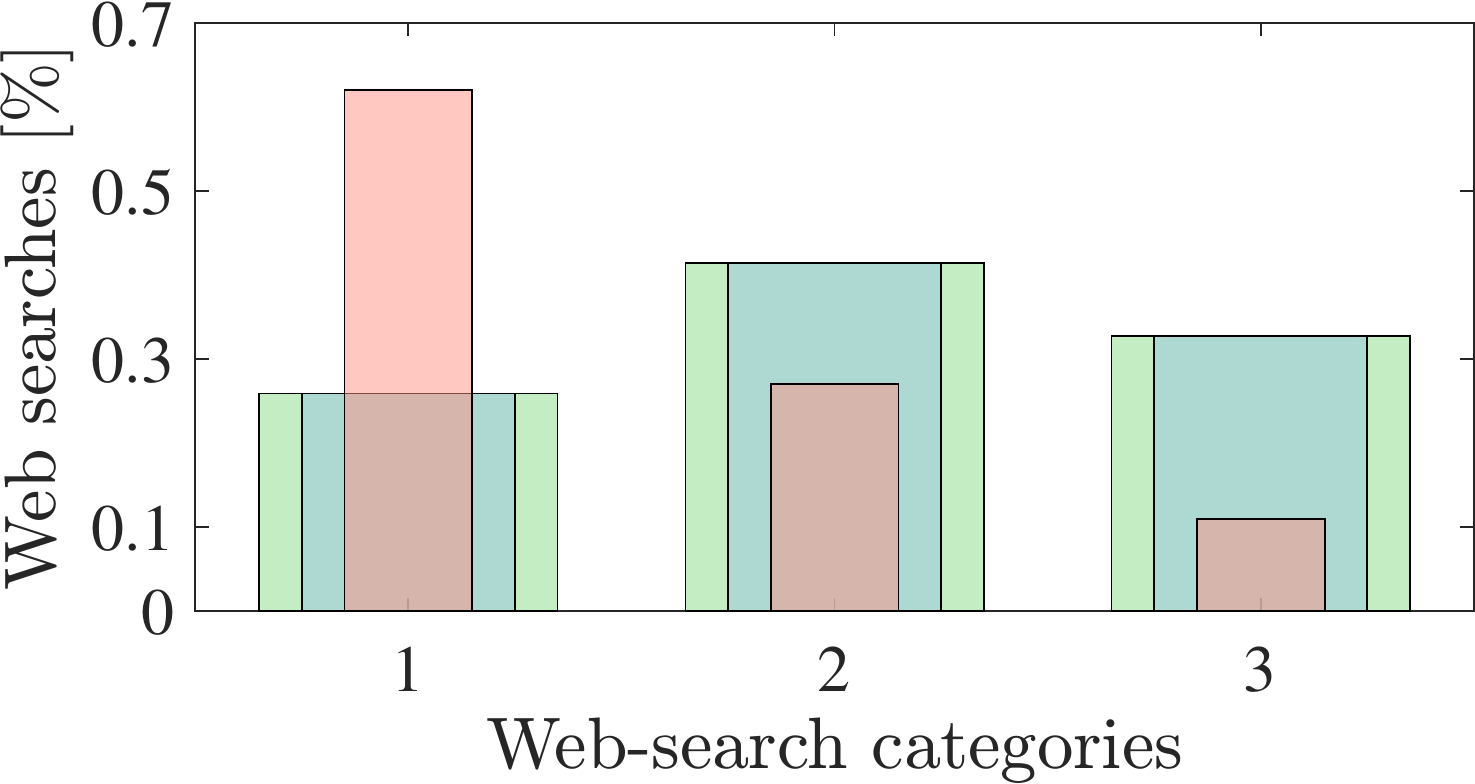}%
\label{}}\hfill
\subfigure[$\mu=\mu_1 \simeq \$0.7948$, $\cR(\mu)\simeq 0.0942$, $\cR(\mu)/\cR(\mumax)\simeq 0.4753$, $\delta^*\simeq (0.6011,0,1)$, $t^*\simeq(0.4760, 0.4140, 0.1100)$.]%
{\includegraphics[scale=0.37]{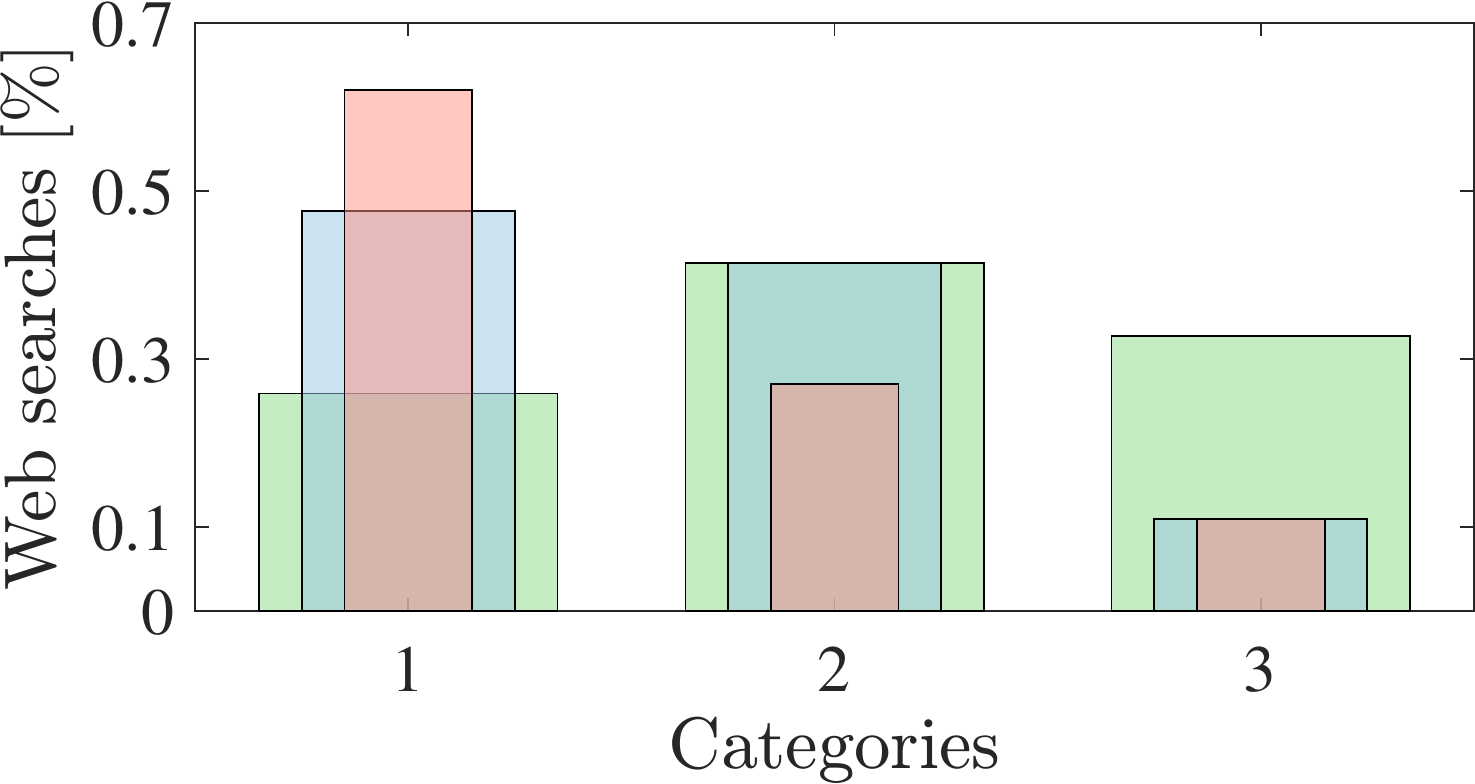}%
\label{}}\hspace*{\fill}
\\
\hspace*{\fill}
\subfigure[$\mu=\$0.8974$, $\cR(\mu)\simeq 0.1358$, $\cR(\mu)/\cR(\mumax)\simeq 0.6853$, $\delta^*\simeq (0.8005,0.4999,1)$, $t^*\simeq(0.5480, 0.3420, 0.1100)$.]%
{\includegraphics[scale=0.37]{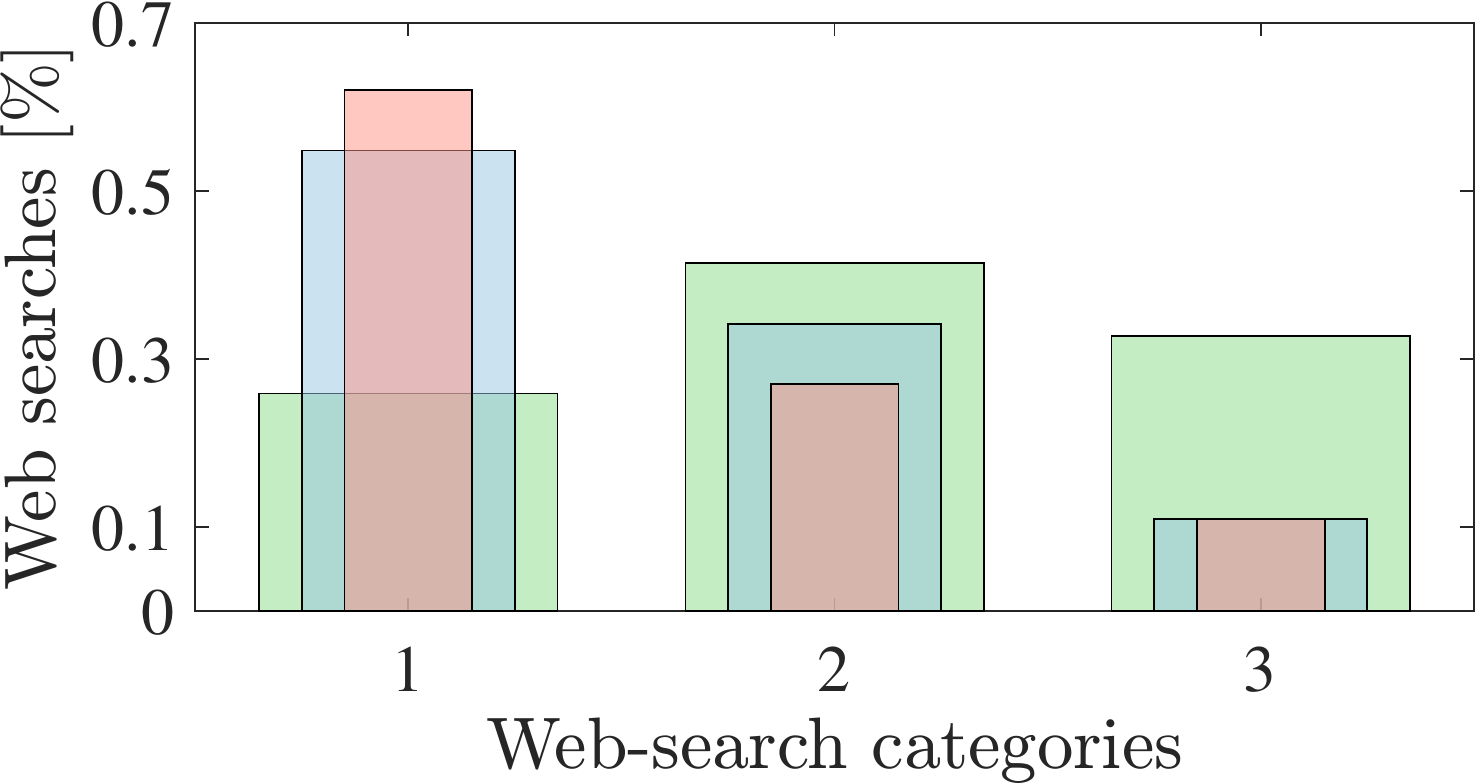}%
\label{}}\hfill
\subfigure[$\mu=\mumax = \$1$, $\cR(\mu)\simeq 0.1981$, $\cR(\mu)/\cR(\mumax)=1$, $\delta^*\simeq (1,1,1)$, $t^*=q$.]%
{\includegraphics[scale=0.37]{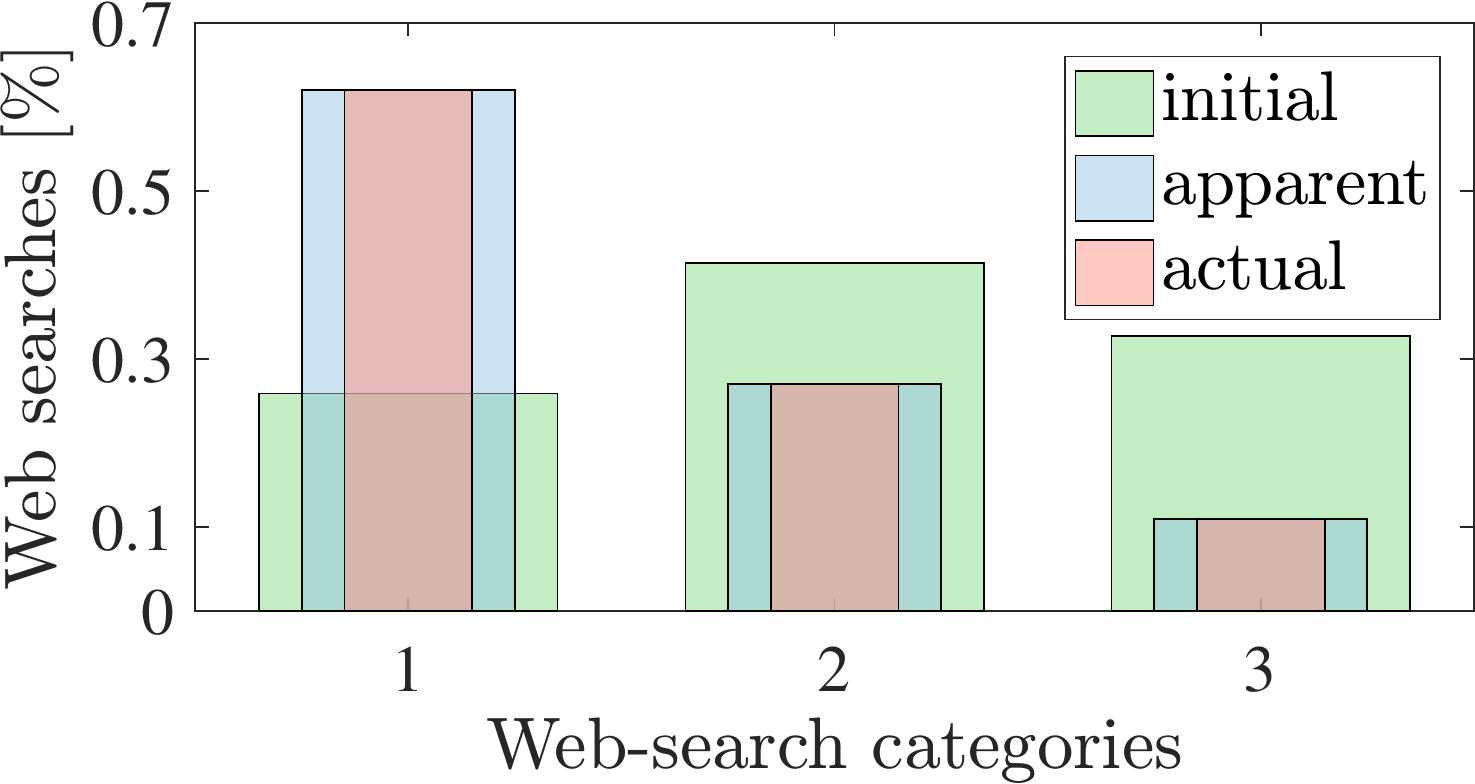}%
\label{}}\hspace*{\fill}%
\caption{Actual, initial and apparent profiles of a particular user for different values of $\mu$.}%
\label{fig:apparentprofiles}%
\end{figure*}

Observe that the range of values of the indexes $k$ and $j$ stated in the theorem corresponds to case~(d) of Lemma~\ref{lem:star}.
The direct application of this lemma in the special case of the SED function leads to the solution $\x_i = \frac{\alpha}{2\,\ad_i} + \frac{w_i\,\beta}{2\,\ad_i^2}$ for $i=k-n,\ldots,n-j+1$, $\x_i = 1$ for $i=n-j+2,\ldots,n$, and $\x_i = 0$ for $i=1,\ldots,k-n-1$.

The system of equations given by $\sum_i \ad_i \x_i=0$ and $\sum_i w_i \x_i = \mu$ has a unique solution since
$\cA_1=0$ and $\ad_i\neq0$ for all $i=1,\ldots,n$.
Routine calculation gives
\vspace{-0.02cm}
\begin{equation*}
\alpha = -\,\overline{m}_{S(k,j)}\,\beta + \frac{2\,\cA_{n-j+2}}{\cardS - n},
\end{equation*}
\vspace{-0.02cm}
\begin{equation*}
\beta = \frac{2\left(\mu - \cW_{n-j+2} + \cA_{n-j+2}\,\overline{m}_{S(k,j)}\right)}{\left(n-\cardS\right)\sigma^2_{m_{S(k,j)}}}.
\end{equation*}
\vspace{-0.02cm}
By plugging these expressions into $\frac{\alpha}{2\,\ad_i} + \frac{w_i\,\beta}{2\,\ad_i^2}$, we derive the components $i=k-n,\ldots,n-j+1$ of the solution.

It remains to confirm the interval of values of $\mu$ in which this solution is defined.
For this purpose, verify first that $\varphi = \arctan \left(\beta/\alpha\right)$ is a strictly monotonic function of $\mu$.
Then, note that the condition $\varphi_k \leqslant \varphi$ in Lemma~\ref{lem:star}, case~(d), 
becomes
\begin{multline*}
- \frac{1}{m_{k-n+1}} \leqslant  - \frac{1}{\overline{m}_{S(k,j)}} + \frac{\cA_{n-j+2}}{\overline{m}_{S(k,j)}}\,\times \\
\times \left(\frac{\overline{m}_{S(k,j)}}{\sigma^2_{m_{S(k,j)}}}\left(\mu - \cW_{n-j+2} + \cA_{n-j+2}\,\overline{m}_{S(k,j)}\right) + \cA_{n-j+2}\right)^{-1}.
\end{multline*}
After simple algebraic manipulation, and on account of $\mu_{k+1,j}<\mu_{k,j}$ and the monotonicity of $\varphi(\mu)$,
we conclude
\begin{equation*}
\mu\leqslant \cW_{n-j+2} - \cA_{n-j+2}\Big(\overline{m}_{S(k,j)} + \frac{\sigma^2_{m_{S(k,j)}}}{\overline{m}_{S(k,j)} - m_{k-n-1}}\Big).
\end{equation*}
An analogous analysis on the upper bound condition $\varphi < \varphi_{k+1}$ determines the interval of values of $\mu$ where the solution is defined.
\EndProof
Although the above theorem only covers the intervals $\mu_{k+1,j}<\mu_{k,j}$ for $k=n+2,\ldots, 2n$ and $j=1,\ldots,2(n+1)-k$, a number of important, intuitive consequences can be drawn from it.
First and foremost, the components $\x_i$ of the form ${\fp'_i}^{-1}(z_i^{\oT} \gamma)$ are linear with the ratio $\frac{v_{i,S(k,j)}}{\ad_i}$,
 exactly as Theorem~\ref{thr:casen3} showed for $n=3$, which means that the optimal strategy follows the same principle described in Sec.~\ref{sec:Theory:CaseN2}. On the other hand, the coincidence of these two results suggests a similar behavior of the solution in a general case.

Another immediate consequence of Theorem~\ref{thr:star} is the role of the money thresholds. In particular,
we identify $\mu_{k,j}$ as the money (paid by a data broker) beyond which the components of $\x_i$ for $i=k-n,\ldots,n$ are all positive.
Conceptually, we may establish an interesting connection between these thresholds and the hyperplanes that determine the solutions space on the $\alpha$-$\beta$ plane.
Lastly, although it has not been proved by Theorem~\ref{thr:star}, we immediately check the quadratic dependence of the trade-off on $\mu$, as shown also in Theorem~\ref{thr:casen3} for $n=3$.

\section{Simple, Conceptual Example}
\label{sec:Examples}
\noindent
In this section, we present a numerical example that illustrates the theoretical analysis conducted in the previous section.
For simplicity, we shall assume the SED as privacy function.

In this example, we consider a user who wishes to sell their Google search profile to one of the new data-broker companies mentioned in Sec.~\ref{sec:Introduction}.
We represent their profile across $n = 3$ categories, namely, ``health'', ``others'' and ``religion'',
as we assume they are concerned mainly with those search categories related to health and religion, whereas the rest of searches are not sensitive to them.  
We suppose that the user's search profile is
\begin{equation*}
q=(0.620, 0.270, 0.110),
\end{equation*}
the initial distribution is
\begin{equation*}
p=(0.259, 0.414, 0.327),
\end{equation*}
and the normalized category rates are
\begin{equation*}
w=(0.404, 0.044, 0.552).
\end{equation*}
The choice of the initial profile and the category rates above may be interpreted from the perspective of a user who hypothetically wants to hide an excessive interest in health-related issues and, more importantly to them, wishes to conceal a lack of interest in religious topics. This is captured by the large differences between $q_1$ and $p_1$ on the one hand, and $q_3$ and $p_3$ on the other, and by the fact that $w_3 > w_1$.

First, we note that $q$ and $p$ satisfy the assumptions~\eqref{eqn:PositivityAssumption} and~\eqref{eqn:NoEqual}, and that $m_1\geqslant m_2 \geqslant m_3$. Also, we verify that $w_{2} \leqslant \ad_{2}\,\overline{m}_{2}$, which, on account of Theorem~\ref{thr:casen3}, implies that the optimal strategy has just two positive components within $\mu \in[0,\mu_1]$, in particular, the categories 1 and 3. Precisely, from Sec.~\ref{sec:Theory:CaseN2}, we easily obtain this money threshold $\mu_1 \simeq \$0.7948$.

From Theorem~\ref{thr:casen3}, we also know that the optimal percentage of disclosure is proportional to the relative coefficient of variation of the ratio $w_i/\ad_i$, which in our example yields
\begin{equation*}
\left(\frac{v_{i,2}}{\ad_i}\right)_i\simeq (1.513, -0.842, 2.516).
\end{equation*}
Accordingly, for $\mu \in[0,\mu_1]$ we expect higher disclosures for category 3, ``religion'', than for category 1, ``health''.
This is illustrated in Fig.~\ref{fig:apparentprofiles}(b), where we plot the actual, initial and apparent profiles for the extreme case $\mu=\mu_1$.
In this figure, we observe that the optimal strategy suggests revealing the user's actual interest completely in category 3. For that economic reward, which accounts for roughly 79.48\% of $\mumax$, interestingly the user sees how their privacy is reduced ``just'' 47.53\%.
Remarkably enough, this unbalanced yet desirable effect is even more pronounced for smaller rewards. For instance, for $\mu=\$0.01$, we note that the increase in privacy risk is only 0.0015\% of the final privacy risk $\cR(\mumax) \simeq 0.1981$.

\begin{figure}[tb]
\centering
\includegraphics[width=\columnwidth]{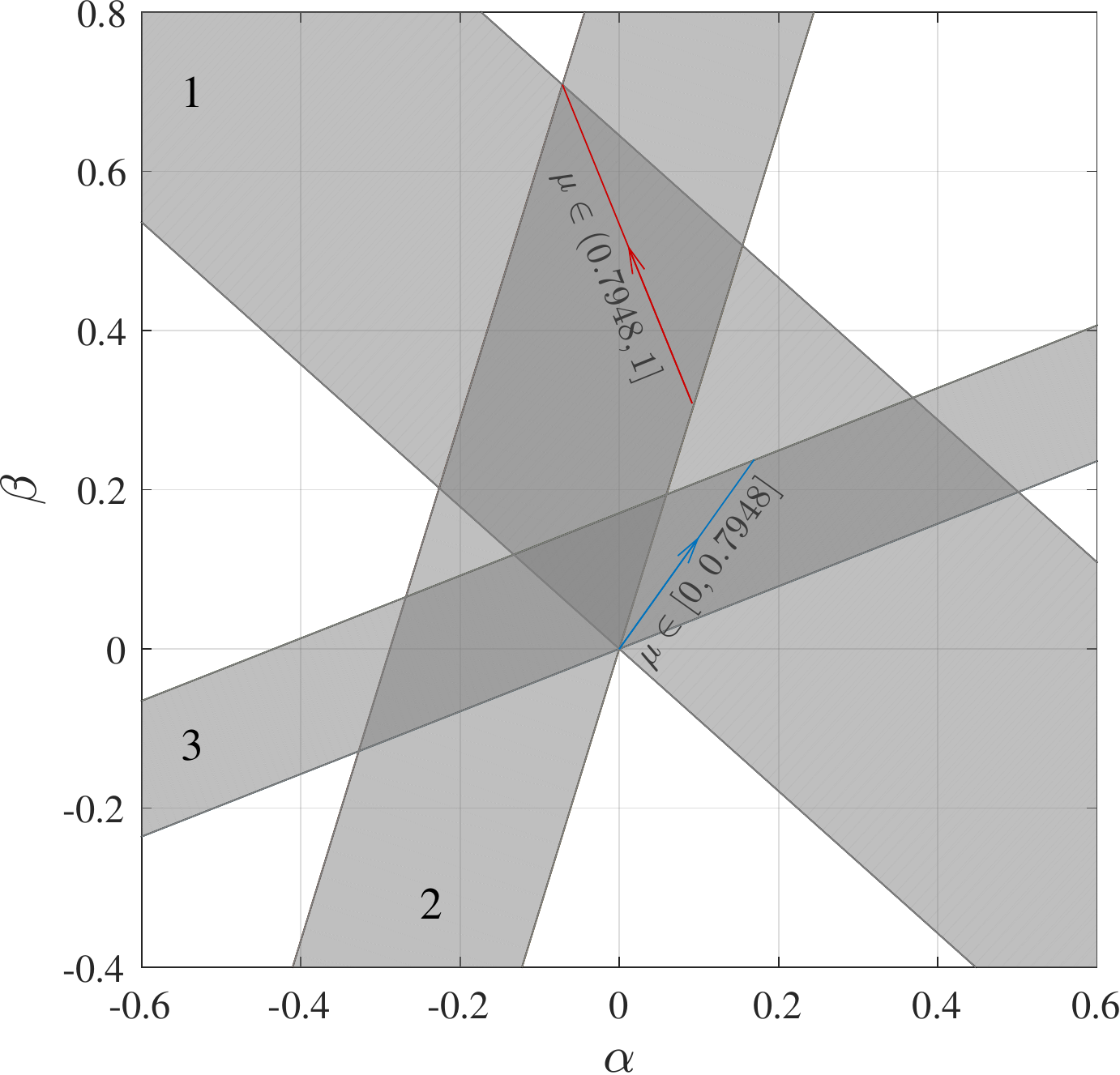}
\caption{Slabs layout on the $\alpha$-$\beta$ plane for the example considered in Sec.~\ref{sec:Examples}. The line segments plotted in blue and red show the dependence of the parameter $\gamma$ on $\mu$.}
\label{fig:example:slabs}
\end{figure}

Recall that $\gamma$ is the parameter that configures the specific point of operation within the $\alpha$-$\beta$ plane in Lemma~\ref{lem:genericproblem}, and thus the specific form (i.e., either 0, 1 or ${\fp'_i}^{-1}(z_i^{\oT} \gamma)$) of each of the components of the optimal disclosure strategy.
In the interval of values $[0,\mu_1]$, the parameter $\gamma$ lies in the closure of halfspaces 1 and 3, as we show in Fig.~\ref{fig:example:slabs}.
An interesting observation that arises from this figure is, precisely, the correspondence between this parameter and $\mu$, and how the latter (obviously together with $q$, $p$ and $w$) determines the former through the primal equality conditions $\sum_i \ad_i\,\x_i = 0$ and $\sum_i w_i \x_i =\mu$.
In particular, we observe that
as $\mu$ increases, $\gamma$ draws a straight line from the lower hyperplane 3 to the upper hyperplane 3, which helps us illustrate how economic rewards are mapped to the $\alpha$-$\beta$ plane.
In addition, because we contemplate the SED function as privacy measure, we appreciate that the three lower hyperplanes intersect at $(0,0)$, as stated in Corollary~\ref{col:specificorigin}.

\begin{figure}[tb]
\centering
\includegraphics[width=\columnwidth]{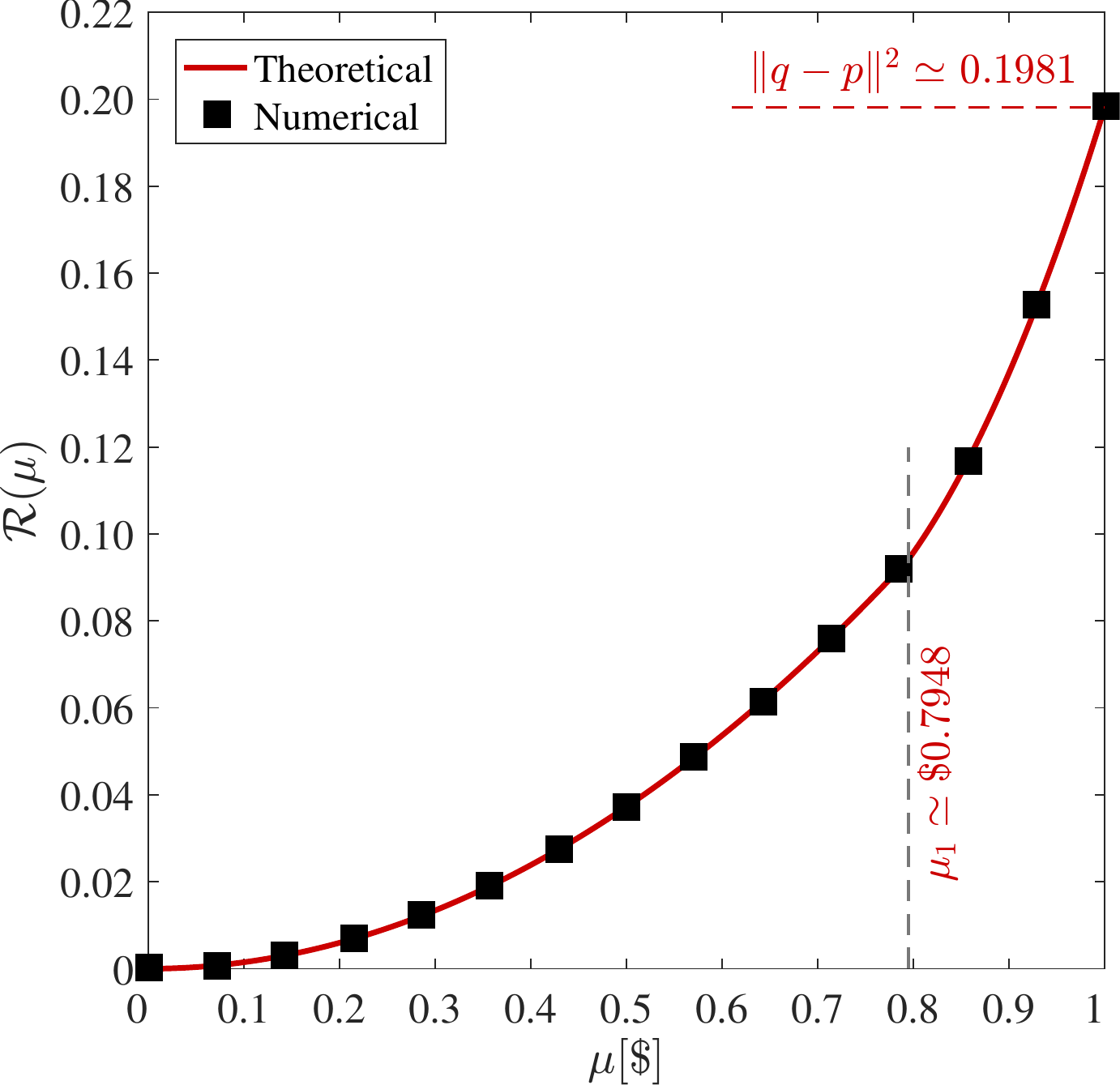}
\caption{Optimal trade-off between privacy and money, the former measured as the SED between the apparent and the initial profiles.}
\label{fig:example:tradeoff}
\end{figure}

To compute the solution to~\eqref{eqn:PrivacyMoneyFunction} for $\mu>\mu_1$, we follow the methodology of the proof of Theorem~\ref{thr:casen3}.
First, we check that the only condition consistent with $\mu_1<\mu<\mumax$ is that $0<\x_1,\x_2<1$ and $\x_3=1$.
We verify this by noting that, when $\x_2=0$, the system of equations given by the above two primal equality conditions is inconsistent.
Then, we notice that, if $0<\x_2<1$, these conditions lead to the following system of equations,
\begin{equation*}
\begin{bmatrix}
1                          & \overline{m}_3 \\
\overline{m}_3 & \frac{1}{2}\sum_{i=1}^2 m_i^2
\end{bmatrix}\,
\begin{bmatrix}
\alpha\\
\beta
\end{bmatrix} =
\begin{bmatrix}
-\ad_3\\
\mu - w_3
\end{bmatrix},
 \end{equation*}
which has a unique solution,
$$(\alpha, \beta)\simeq \left(- 0.8017\,\mu + 0.7303, 1.9708\,\mu - 1.2619\right).$$
From this solution, it is immediate to obtain the optimal strategy $\delta_1^* (\mu)\simeq  1.9444\,\mu - 0.9445$ and $\delta_2^* (\mu)\simeq 4.8746\,\mu - 3.8746$.
Following an analogous procedure, we find that its interval of validity is $(\mu_1,\mumax]$, where we note that $\mumax=\$1$.

From the expressions of $\x_1$ and $\x_2$ above, we observe that the optimal strategy unveils the actual interest values of both categories only when $\mu=\mumax$, in which case $t=q$. This is plotted in Fig.~\ref{fig:apparentprofiles}(d).
An intermediate value of $\mu$ is assumed in Fig.~\ref{fig:apparentprofiles}(c) that allows us to show the distinct rates of disclosure for the category 1 between the cases $\mu\in[0,\mu_1]$ and $\mu\in(\mu_1,1]$. In particular, the rate of profile disclosure is 0.7560 for the former interval, whereas the optimal strategy recommends a significantly larger rate for the latter interval (1.9444).
The interval of operation $(\mu_1,1]$, on the other hand, places $\gamma$ on the intersection between slabs 1 and 2. Fig.~\ref{fig:example:slabs} shows this and how $\gamma$ approaches to the intersection between the upper hyperplanes 1 and 2 as $\mu$ gets close to $\$1$.

Finally, Fig.~\ref{fig:example:tradeoff} depicts the privacy-money function $\cR(\mu)$, which characterizes the optimal exchange of money for privacy for the user in question. The results have been computed theoretically, as indicated above, and numerically,
and confirm the monotonicity and convexity of the optimal trade-off, proved in Theorems~\ref{thr:Monotonicity} and~\ref{thr:Convexity}.

\section{Related Work}
\label{sec:SotA}
\noindent
To the best of our knowledge, this work is the first to mathematically investigate a hard-privacy mechanism by which users themselves ---without the need of any intermediary entity--- can sell profile information and achieve serviceable points of operation within the optimal trade-off between disclosure risk and economic reward.
As we shall elaborate next,
quite a few works have investigated the general problem of sharing private data in exchange for an economic compensation.
Nevertheless, they tackle different, albeit related, aspects of this problem:
some assume an interactive, query-response data release model~\cite{Ghosh11EC,Li13ICDT,Aperjis12FS,Riederer11HOTNET,Dandekar14JEC}
and aim at assigning prices to noisy query answers~\cite{Ghosh11EC,Li13ICDT,Riederer11HOTNET};
most of them assume distinct purchasing models where data buyers are not be interested in the private data of any particular user, but in aggregate statistics about a large population of users~\cite{Ghosh11EC,Li13ICDT,Aperjis12FS,Riederer11HOTNET,Dandekar14JEC};
the majority of the proposals limit their analysis to differential privacy~\cite{Dwork06A} as measure of privacy~\cite{Ghosh11EC,Li13ICDT,Riederer11HOTNET,Dandekar14JEC};
and some rely on a soft-privacy model whereby users entrust an external entity or trusted third party
to safeguard and sell their data~\cite{Ghosh11EC,Li13ICDT,Aperjis12FS,Dandekar14JEC}.
In this section we briefly examine several of those proposals, bearing in mind that
none of them are user-centric and consider that data owners can sell their profile data directly to brokers.

The study of the monetization of private data was first investigated formally in~\cite{Ghosh11EC}.
The authors tackled the particular problem of pricing private data~\cite{Roth12SIGECOM} in
a purchasing model composed of data owners, who contribute their private data;
a data purchaser, which sends aggregate queries over many owners' data; and a data broker, which is entrusted those data, replies and charges the buyer, and ultimately compensates the owners.
Accordingly, the problem consists in assigning prices to noisy answers, as a function of their accuracy, and how to distribute the money among data owners who deserve compensation for the privacy loss incurred.
The operation of the monetization protocols may be described conceptually as follows: in response to a query, the data broker computes the true query answer, but adds random noise to protect the data owners' privacy. By adding perturbation to the query answer, the price can be lowered so that the more perturbation is introduced, the lower the price is charged. The data buyer may indicate to this end how much precision it is willing to pay for when issuing the query, similarly to our data-purchasing model where we assume buyers start bidding before any disclosure is made.

Various extensions and enhancements were introduced later in~\cite{Li13ICDT,Fleischer12EC,Ligett12ICINE,Roth12EC,Dandekar11arXiv}.
The most relevant is~\cite{Li13ICDT},
which also capitalizes on differential privacy to quantify privacy,
but differs in that it permits several queries and does not require that the minimum compensation users want to receive be public information (as we assume in this work).
This approach, however, cannot be applied to the problem at hand since it relies on a distinct purchasing model where data buyers are not concerned with a single user's data, but aim to obtain aggregate statistics about a population through an interactive, query-response database.
This is in stark contrast to our approach, which assumes buyers are interested in purchasing profile data of particular users, for example, to provide personalized, tailored services such as behavioral advertising~\cite{Goldfarb11COMACM}.

Another related work is~\cite{Aperjis12FS}, which considers a rather simple mechanism to regulate the exchange of money for private data.
The proposed setting permits a buyer to select the number of data owners to be involved in the response to its query.
The mechanism is based on the assumption that a significant portion of data owners show risk-averse behaviors~\cite{Holt02JAER}.
The operation of the mechanism, however, leaves users little control over their data: a market maker is the one deciding whether to disclose the whole data of an individual or to prevent any access to this information.
Our data-buying model does not consider these two extremes,
but the continuum in between
enabled by a disclosure mechanism designed to attain the optimal privacy-money trade-off.
Finally, \cite{Riederer11HOTNET}~proposes auction mechanisms to sell private information to data aggregators.
But again, the data of a particular user are either completely hidden or fully disclosed, and the compensation is determined by buyers without allowing for users' personal privacy valuations.

\section{Conclusions}
\label{sec:Conclusions}
\noindent
This work examines a mechanism that gives users direct control over the sale of their private data.
The mechanism relies on a variation of the purchasing model proposed by the new broker firms which is in line with the literature of pricing private data.

The objective of this paper is to investigate mathematically the privacy-money trade-off posed by this mechanism.
With this aim,
we formulate a multiobjective optimization problem characterizing the trade-off between profile disclosure on the one hand, and on the other economic reward.
Our theoretical analysis provides a general parametric solution to this problem, which is derived for additively separable, twice differentiable privacy functions, with strictly increasing derivatives.
We find that the optimal disclosure strategy exhibits a maximin form, depends on the inverse of the derivative of a privacy function, and leads to a nondecreasing and convex trade-off.
The particular form of each of the $n$ components of the solution, however, is determined by the specific configuration of $2n$ halfspaces,
which in turn depend on the particular values of $q,p,w,\mu$ and $n$.

To proceed towards an explicit closed-form solution, we study some examples of privacy functions and particular cases of those variables.
Specifically, we derive riveting results for several Bregman divergences, although special attention is given to the SED function.

In our analysis, we verify the existence of an origin of coordinates in the slabs layout that permits us to leverage certain regularities.
For $n\leqslant 3$ and a general configuration of slabs, we show the dependence of the closed-form solution (essentially) on Fano's factor and the intuitive principle behind the optimal strategy, which recommends disclosing a profile mostly in those categories where $\ad_i$ is small and $m_i$ deviates the most from its mean value, compared to its variance.

For arbitrarily large $n$, we investigate a concrete slabs layout that allows us to obtain an explicit closed-form expression of both the solution and trade-off. The configuration of slabs, which we call conical regular, permits parameterizing the solution with polar coordinates.
The optimal strategy is also a piecewise linear function of the same index of dispersion, which may indicate a similar behavior of the solution in a general configuration.
Our findings show that the form attained by each of the components of the solution is determined by a sequence of thresholds,
which we interpret geometrically as lower hyperplanes.
Finally, our formulation and theoretical analysis are illustrated with a numerical example.

\section*{Acknowledgment}
\noindent
This work was partly funded by the European Commission through the project H2020-644024 ``CLARUS'', the Spanish Ministry of Economy, Industry and Competitiveness (MINECO) through the project TIN2016-80250-R ``Sec-MCloud'', as well as by the Government of Catalonia under grant 2014 SGR 00537.
J. Parra-Arnau is the recipient of a Juan de la Cierva postdoctoral fellowship, FJCI-2014-19703, from the MINECO.

\end{document}